\documentclass[useAMS, usenatbib]{mn2e}
\usepackage{lscape}
\usepackage{rotating}
\usepackage{amssymb}
\usepackage{amsmath}
\usepackage{float}
\usepackage{hyperref}
\hypersetup{colorlinks = true, linkcolor = blue, urlcolor=blue, citecolor=blue} 
\usepackage[normalem]{ulem}
\newcommand{\GG}[1]{}
\usepackage{tabularx}
\usepackage{rotating, booktabs}

\def\msun{\hbox{M$_\odot$}}

\def\t4{\hbox{t$_{\rm 4}$}}

\def\cm3{\hbox{cm$^{-3}$}}

\usepackage{graphicx}
\title[Coeval MPs in NGC 1978]
{The search for multiple populations in Magellanic Cloud Clusters IV: Coeval multiple stellar populations in the 
young star cluster NGC~1978}
\author[Martocchia et al.] {S. Martocchia$^{1}$, F. Niederhofer$^{2}$, E. Dalessandro$^{3}$, N. Bastian$^{1}$, N. Kacharov$^{4}$, 
\newauthor C. Usher$^{1}$, I. Cabrera-Ziri$^{5}$\thanks{Hubble Fellow.}, C. Lardo$^{6}$, S. Cassisi$^{7}$, D. Geisler$^{8}$, M. Hilker$^{9}$, 
\newauthor   K. Hollyhead$^{10}$, V. Kozhurina-Platais$^{11}$, S. Larsen$^{12}$,  D. Mackey$^{13}$, 
\newauthor  A. Mucciarelli$^{3, 14}$, I. Platais$^{15}$, M. Salaris$^{1}$ \\
$^{1}$Astrophysics Research Institute, Liverpool John Moores University, 146 Brownlow Hill, Liverpool L3 5RF, UK\\
$^{2}$Leibniz-Institut f\"ur Astrophysik Potsdam, An der Sternwarte 16, Potsdam 14482, Germany\\
$^{3}$INAF-Osservatorio di Astrofisica \& Scienza dello Spazio, via Gobetti 93/3, I-40129, Bologna, Italy\\
$^{4}$Max-Planck-Institut f\"ur Astronomie, K\"onigstuhl 17, D-69117 Heidelberg, Germany\\
$^{5}$Harvard-Smithsonian Center for Astrophysics, 60 Garden Street, Cambridge, MA 02138, USA\\
$^{6}$Laboratoire d'astrophysique, École Polytechnique Fédérale de Lausanne (EPFL), Observatoire, 1290, Versoix, Switzerland\\
$^{7}$INAF, Osservatorio Astronomico d'Abruzzo, sn. 64100 Teramo, Italy\\
$^{8}$Departamento de Astronomia, Universidad de Concepcion, Casilla 160-C, Chile\\
$^{9}$European Southern Observatory, Karl-Schwarzschild-Stra\ss e 2, D-85748 Garching bei M\"unchen, Germany\\
$^{10}$Department of Astronomy,Oscar Klein Centre,Stockholm University, AlbaNova, Stockholm SE-10691,Sweden\\
$^{11}$Space Telescope Science Institute, 3700 San Martin Drive, Baltimore, MD 21218, USA\\
$^{12}$Department of Astrophysics/IMAPP, Radboud University, P.O. Box 9010, 6500 GL Nijmegen, The Netherlands\\
$^{13}$Research School of Astronomy and Astrophysics, Australian National University, Canberra, ACT 2611, Australia\\
$^{14}$Dipartimento di Fisica \& Astronomia, Universit\` a degli Studi di Bologna, via Gobetti 93/2, I-40129, Bologna, Italy\\
$^{15}$Department of Physics and Astronomy, Johns Hopkins University, 3400 North Charles Street, Baltimore, MD 21218, USA\\
}
\date{Accepted. Received ; in original form.}
\pagerange{\pageref{firstpage}--\pageref{lastpage}}
\pubyear{2017}

\begin{document}
\maketitle
\label{firstpage}

\begin{abstract}We have recently shown that the $\sim2$ Gyr old Large Magellanic Cloud star cluster NGC 1978 hosts multiple populations in terms of star-to-star abundance variations in [N/Fe]. These can be seen as a splitting or spread in the sub-giant and red giant branches (SGB and RGB) when certain photometric filter combinations are used.  
Due to its relative youth, NGC 1978 can be used to place stringent limits on whether multiple bursts of star-formation have taken place within the cluster, as predicted by some models for the origin of multiple populations.  
We carry out two distinct analyses to test whether multiple star-formation epochs have occurred within NGC 1978.  
First, we use UV CMDs to select stars from the first and second population along the SGB, and then compare their positions in optical CMDs, where the morphology is dominantly controlled by age as opposed to multiple population effects.
We find that the two populations are indistinguishable, with age differences of $1\pm20$ Myr between them.  This is in tension with predictions from the AGB scenario for the origin of multiple populations.
Second, we estimate the broadness of the main sequence turnoff (MSTO) of NGC 1978 and we report that it is consistent with the observational errors. We find an upper limit of $\sim$65 Myr on the age spread in the MSTO of NGC 1978. This finding is in conflict with the age spread scenario as origin of the extendend MSTO in intermediate age clusters, while it fully supports predictions from the stellar rotation model. 
\end{abstract}

\section{Introduction}
\label{sec:intro}

Our understanding of globular clusters (GCs) and their formation has undergone a radical change in the past decades.
GCs are found to host subpopulations of stars with distinctive light element abundance patterns. Such chemical multiple populations (MPs) manifest in the form of star-to-star anti-correlated light element abundance variations, such as enhanced N and Na, along with depleted C and O \citep{cannon98, carretta09}.

Due to the type of elements involved in the variations, chemical anomalies were thought to be caused by self-enrichment in elements that originate from high temperature H burning in the interiors of stars, requiring multiple epochs of star formation within the same cluster. 
The general picture is that a first population of stars (FP) forms and the gas not used in star formation is expelled. Material processed by the FP stars  
collects in the centre of the cluster and then mixes up with material with ``pristine'' composition from the outskirts of the cluster, which is re-accreted.
A second population (SP) is eventually formed by a second burst of star formation. 
Depending on which ``polluter'' is advocated, i.e. the source of enrichment of the chemical anomalies, such multiple generational theories predict
age spreads from a few Myr (massive and super-massive stars, e.g. \citealt{decressin07,demink09}, \citealt{denissenkov14}, Gieles et al. in prep.) to 30-200 Myr (asymptotic giant branch, AGB stars, e.g. \citealt{dercole08, conroyspergel11}).

An immediate test can be implemented to verify the nature of the polluters, by estimating the age difference among the subpopulations in a cluster. 
This has been attempted in studies of ancient GCs (e.g., \citealt{marino12, nardiello15}) but due to the old ages, only upper limits of $\sim$ 200 Myr between the populations have been achieved.  If such an experiment could be carried out on younger clusters, more stringent limits could be placed.

We have recently undertaken a Hubble Space Telescope (HST) photometric survey of 9 massive ($\gtrsim 10^5$ \msun) star clusters in the Magellanic clouds (MCs) in order to study the behaviour of the MP phenomenon in an unexplored region of the parameter space, i.e. younger ages. To date, we found that all clusters older than 2 Gyr show MPs, while all clusters younger than this age do not \citep{paperI,paperII,paperIII,martocchia17b}. 
One of the most remarkable results from our survey is that NGC 1978, a massive (2-4 $\times 10^5$ \msun, \citealt{westerlund97}) and relatively young ($\sim$ 2 Gyr old, \citealt{mucciarelli07}) cluster in the Large Magellanic Cloud (LMC),
shows evidence for MPs in its red giant branch (RGB, \citealt{martocchia17b}). To date, this is the youngest cluster where the presence of chemical anomalies have been detected. 
This can be constrasted with NGC 419 which has very similar properties (mass, radius), but it is $\sim$ 500 Myr younger and does not appear to host MPs. 

Another open question regarding the MP phenomenon is the connection between chemical anomalies and the presence of an extended main sequence turnoff (eMSTO) in young MCs clusters. The eMSTO feature is observed in young ($>20$ Myr, e.g. \citealt{milone15,bastian16}) and intermediate-age ($<2$ Gyr, e.g. \citealt{mackey08,milone09}) massive clusters.
Many authors also refer to the term ``multiple populations'' as the presence of broadenings and/or splittings in the turnoff (TO) and main sequence (MS) regions in colour magnitude diagrams (CMDs) of young/intermediate-age clusters. However, no star-to-star chemical abundance variations appear to be associated with this phenomenon \citep{mucciarelli08,mucciarelli12,mucciarelli14}, thus we will use the term MPs to only refer to the feature involving chemical anomalies, i.e. spreads and splitting in the RGB and/or SGB.

It was originally thought that the eMSTO was due to age spreads of several hundreds of Myr caused by multiple star forming events (e.g. \citealt{goudfrooij14}). Such a scenario has been demonstrated to have several caveats.  
Among these, massive clusters should be forming stars for the first 10s to 100s Myr of their lives, while no clusters with ages of $\sim 10$ Myr or older have been found to host current star formation events \citep{bastian13,cabrera16}. Additionally, there is a clear correlation between the inferred age spread and the age of the cluster, thus suggesting a stellar evolutionary effect is the cause \citep{niederhofer15}.

Alternatively, recent works have shown that the eMSTO is likely due to a single age population with a range of stellar rotation rates \citep{bastiandemink09,niederhofer15,brandt15}. In the stellar rotation paradigm, clusters older than 2 Gyr would not be expected to host eMSTOs, since stars on the TO in 2 Gyr old clusters are able to develop convective cores, and subsequently host magnetic fields which can brake the star, i.e. no rapid rotators would be expected.

In this work, we take advantage of the unique characteristics of NGC 1978 to place stringent limits on any internal age dispersion. 
First, we will use the subgiant branch (SGB) of NGC 1978 to search for age differences between the different populations in the cluster. Then, we explore the morphology of the MSTO of NGC 1978, showing that this is consistent with a single aged population.

This paper is organised as follows: \S \ref{sec:observations} outlines the photometric reduction procedures for NGC 1978 and the description of the isochrones adopted for the analysis. 
In \S \ref{sec:sgb} we present the analysis of the SGB of NGC 1978, while in \S \ref{sec:sfh} we report on the analysis of the MSTO. We finally discuss our results and conclude in \S \ref{sec:discussion}.

\section{Observations and adopted isochrones}
\label{sec:observations}

HST imaging of NGC 1978 was obtained from our ongoing HST survey (GO-14069 program, PI: N. Bastian, see \citealt{paperI} for more details)\footnote{The photometric catalogues are available from the authors upon request.}.
The new filters are F336W, F343N and F438W (WFC3/UVIS camera), introduced because of their sensitivity to MPs, which are combined with archival observations in F555W and F814W filters (ACS/WFC, GO-9891, PI: G. Gilmore).

Photometry and data reduction procedures for NGC 1978 were extensively reported in \cite{martocchia17b}, and we refer the interested reader to this paper for more details about the photometric catalogue. 
We will also refer to F336W as $U$, F343N as $Un$, F438W as $B$, F555W as $V$ and F814W as $I$ hereafter for simplicity, unless stated otherwise. We will rather report the original 
filter names in the figures. 

In order to translate observational quantities (in units of mag) into age differences in Myr, we used the BaSTI models (``A Bag of Stellar Tracks and Isochrones", \citealt{pietrinferni04}). This choice allowed to properly account for the effects of core convective overshooting during the central H-burning stage. We note that the overshooting efficiency $\Lambda_{\rm ov}$ is commonly parametrised as a fraction of the pressure scale height ($H_P$). In case of the BaSTI database the following dependence of $\Lambda_{\rm ov}$ as a function of the mass: (i) $\Lambda_{\rm ov} = 0.2 H_p$ for masses larger than 1.7 \msun , (ii) $\Lambda_{\rm ov} = (M/$\msun $-0.9)H_p/4$ for stars between 1.1-1.7 \msun , and (iii) $\Lambda_{\rm ov}=0 H_p$ for stars less massive than 1.1 \msun.
Stellar masses for a 2 Gyr old cluster are $\sim$ 1.45 \msun \, at the MSTO and $\sim$ 1.47-1.49 \msun \, on the SGB.
BaSTI isochrones spaced by 20 Myr in age were specifically calculated for this work.

\begin{figure}
\centering
\includegraphics[scale=0.7]{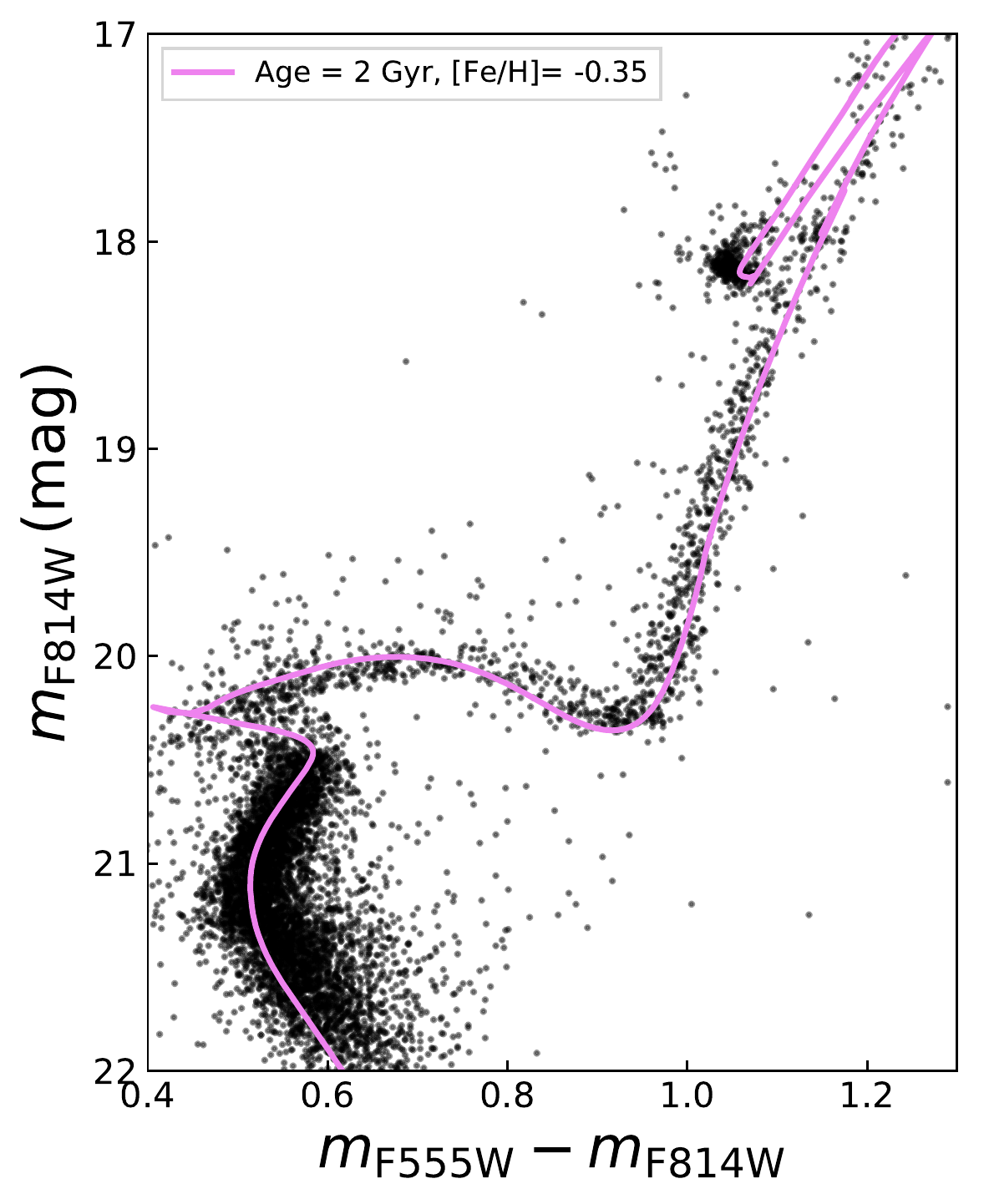}
\caption{$V-I$ vs. $I$ CMD of NGC 1978. The violet curve indicates the BaSTI isochrone for values of age $=2$ Gyr and metallicity [Fe/H]$=-0.35$ dex. The adopted distance modulus and extinction
values are DM $=$ 18.55 mag and $A_{V}=0.16$ mag, respectively.} 
\label{fig:iso}
\end{figure}

Figure \ref{fig:iso} shows the $V-I$ vs. $I$ CMD of NGC 1978 with a BaSTI isochrone superimposed as a solid violet line. The chosen parameters are the following: (i) age $= 2$ Gyr, (ii) metallicity [Fe/H] $= -0.35$ dex, and (iii) distance modulus DM $=18.55$ mag, (iv) extinction value $A_V = 0.16$ mag. The isochrone in Fig. \ref{fig:iso} nicely reproduces the shape of the $V-I$ vs. $I$ CMD in all its evolutionary stages. 
The adopted metallicity is consistent with results by \cite{ferraro06}, who found an iron content of [Fe/H]$=-0.38$ dex by analysing 11 high-resolution FLAMES spectra of giant stars in NGC 1978.

Additionally, \cite{mucciarelli07} performed an isochrone fitting of the ACS $V-I$ vs. $V$ CMD of NGC 1978 by adopting several sets of isochrones. They measured the age of the cluster to be 1.9$\pm$0.1 Gyr, which is consistent with the age we adopt.

\subsection{Artificial stars test}
\label{subsec:astest}

We performed artificial star (AS) experiments following the method described in \cite{dalessandro15} (see also \citealt{bellazzini02,dalessandro16}) to derive a reliable estimate of the photometric errors. AS were performed for the entire data-set adopted in the present paper. In particular, we note that they are especially critical for both the F555W and F814W bands as only one image is available for these filters.

We generated a catalog of simulated stars with an $I$-band input magnitude (I$_{in}$) extracted from a luminosity function (LF) modeled to reproduce the observed LF in that band and extrapolated beyond the observed limiting magnitude. We then assigned a $U_{in}$, $Un_{in}$, $B_{in}$ and V$_{in}$ magnitudes to each star extracted from the luminosity function, by means of an interpolation along the ridge mean lines that were obtained in different CMDs by averaging over 0.4 mag bins and applying a $2\sigma$ clipping algorithm.

Artificial stars were added to real images (which include also real stars) by using the software {\tt DAOPHOTII/ADDSTAR} \citep{stetson87}.
Then, the photometric analysis was performed using the same reduction strategy and PSF models used for real images (see \citealt{martocchia17b} for details) on both real and simulated stars. In this way, the effect of radial variation of crowding on both completeness and photometric errors is accounted for. Artificial crowding was minimized by placing stars into the images following a regular grid composed by $25\times 25$ pixel cells in which only one artificial star for each run was allowed to lie at a random position within the cell. 
For each run, we simulated in this way $\sim 14,000$ stars. The procedure was repeated several times until a minimum number of 100,000 was added to each ACS (for $V$ and $I$) or WFC3 (for $U_n$, $U$ and $B$) chip. At each run the positions of the simulated stars are randomly changed. After a large number of experiments, stars are uniformly distributed in coordinates. Stars recovered after the AS photometric analysis have values of $U_{out}$, $Un_{out}$, $B_{out}$, V$_{out}$ and I$_{out}$.

\begin{figure*}
\centering
\includegraphics[scale=0.8]{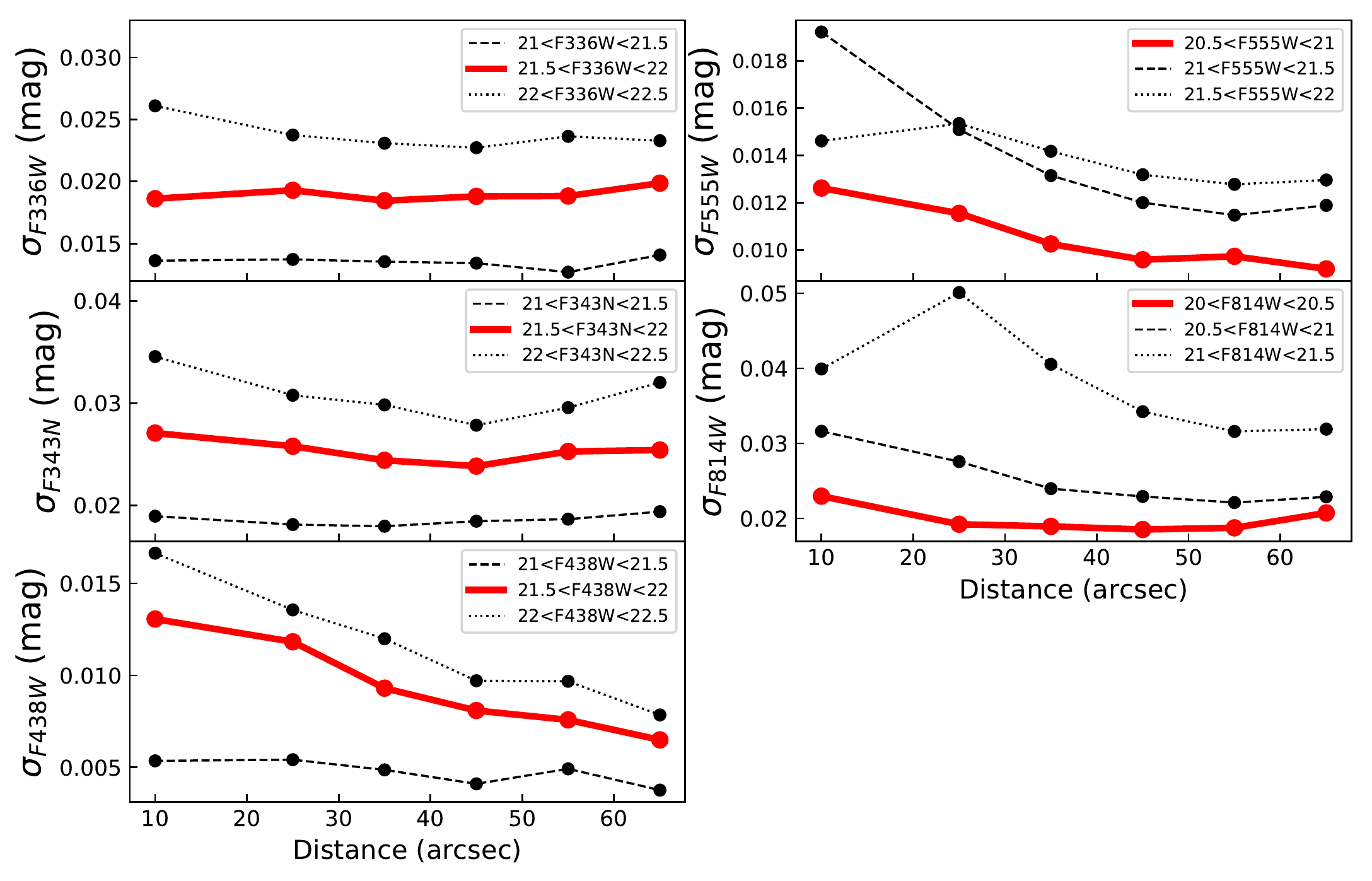}
\caption{Photometric errors for the F336W($U$), F343N($Un$), F438W($B$), F555W($V$), F814W($I$) filter bands as a function of the distance from the cluster centre. This is shown for three bins of magnitude in each panel. The red line indicates the magnitude bin corresponding to the SGB.}
\label{fig:photerr}
\end{figure*}

The AS catalog was then used to derive photometric errors for SGB and MS stars, applied in the following analysis (see Sections \ref{sec:sgb} and \ref{sec:sfh}). Errors were derived computing the r.m.s. of the distributions of simulated stars in the (mag$_{in}$,mag$_{in}$ $-$ mag$_{out}$) diagrams for all available bands in different magnitude bins and after applying the same selections in the photometry quality indicators (sharpness and chi) that were originally applied to the data (see \citealt{martocchia17b}). The distribution of the average errors as a function of the distance from the cluster centre for all the considered bands is shown in Figure \ref{fig:photerr}.  As expected, in the optical images ($B$, $V$ and $I$) where crowding is stronger, errors progressively decrease moving towards the external regions of the cluster as crowding becomes less and less severe. On the contrary we do not observe any significant variation in the UV filters ($U_n$ and $U$).

\begin{figure}
\centering
\includegraphics[scale=0.4]{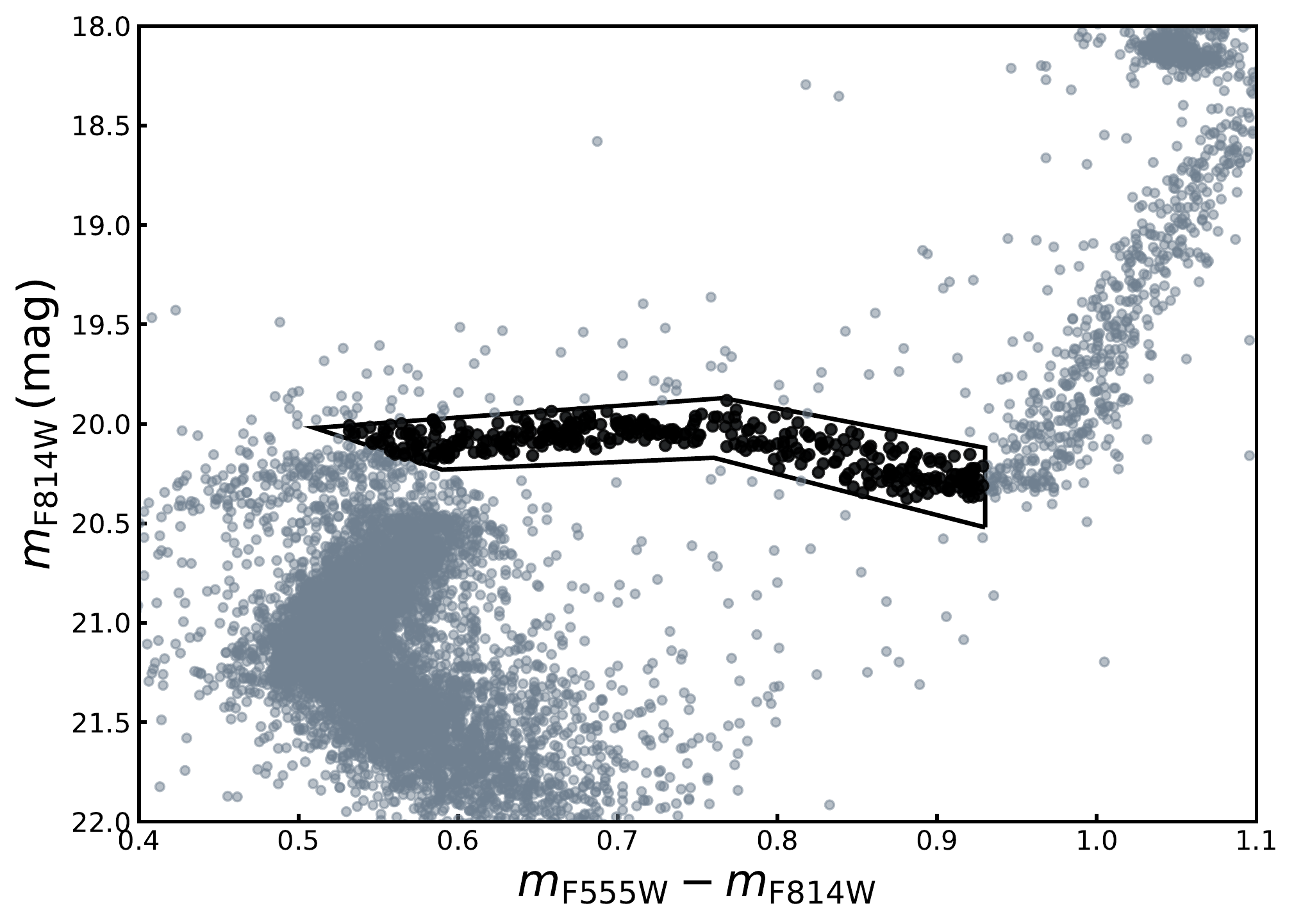}
\caption{$V-I$ vs. $I$ CMD of NGC 1978. The black box indicates the locus of the initial selection of SGB stars which are marked by black filled circles. } 
\label{fig:sgbsel}
\end{figure}

\begin{figure*}
\centering
\includegraphics[scale=0.37]{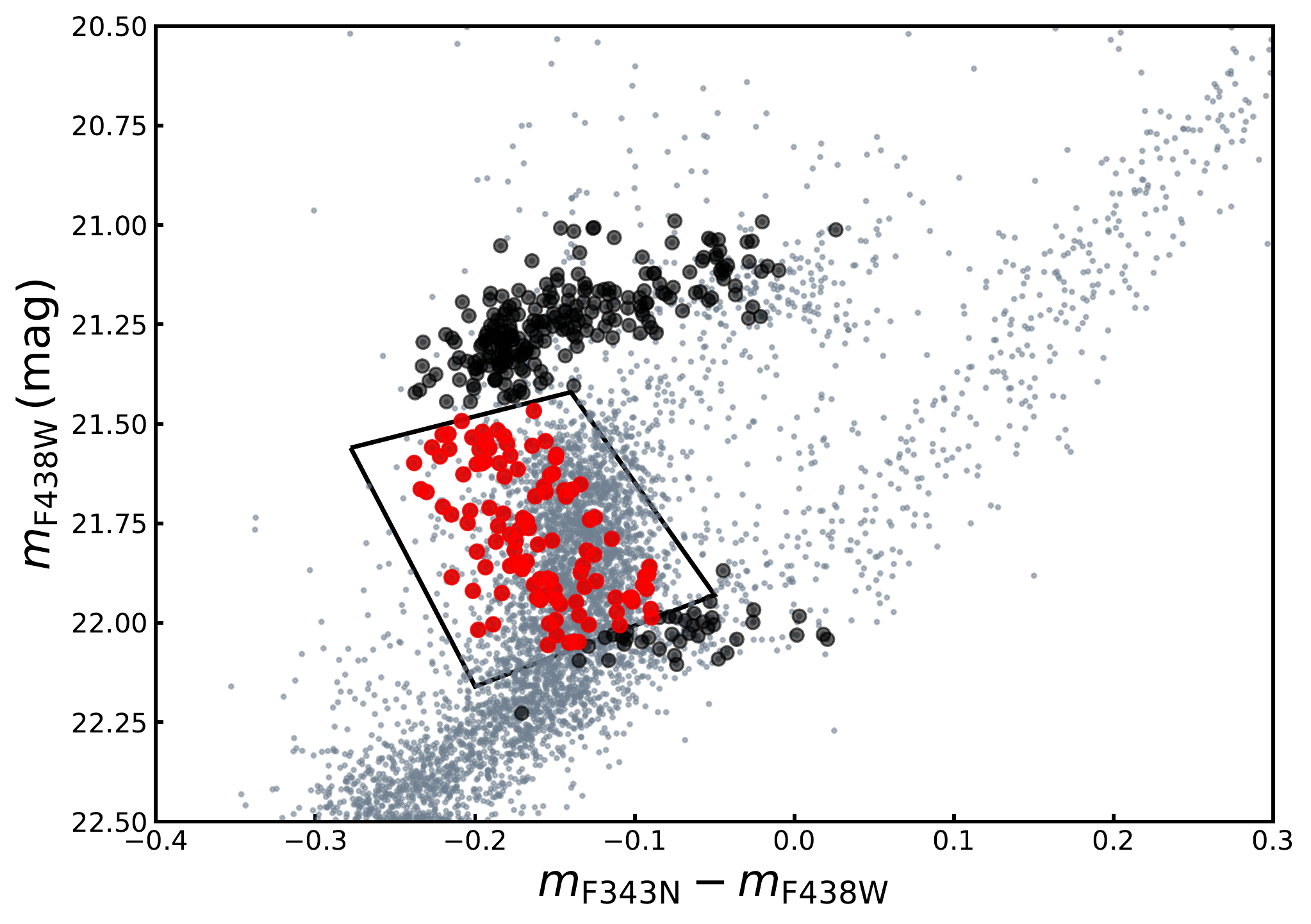}
\includegraphics[scale=0.37]{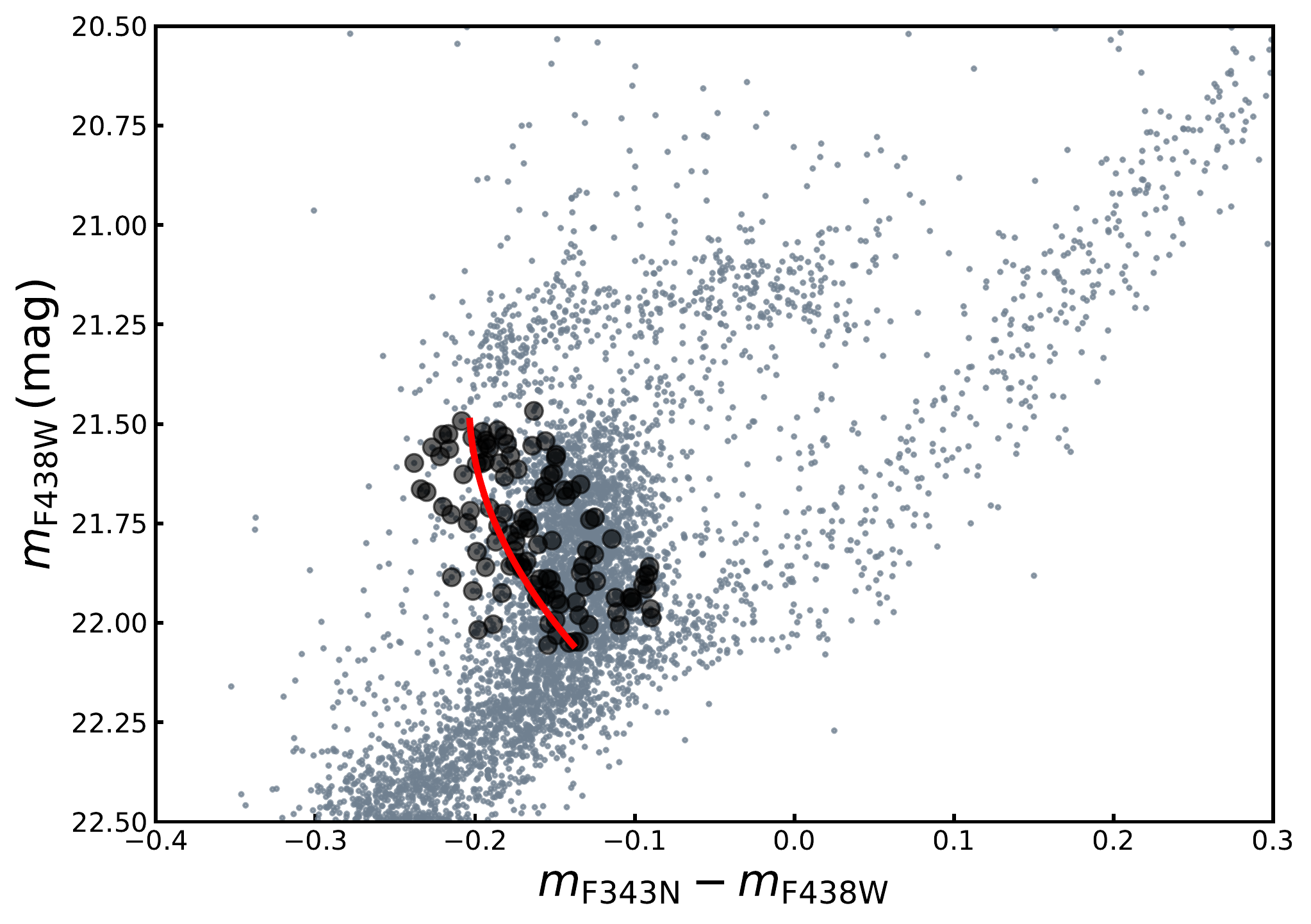}
\caption{$Un-B$ vs $B$ CMDs of NGC 1978. {\it Left Panel:} Red filled circles indicate the final selected SGB stars, while black circles represent the stars that did not pass the final selection.
{\it Right Panel:} Black circles mark the final selected SGB stars. The red solid line indicates the fiducial line defined on the edge of the SGB.}
\label{fig:UnB}
\end{figure*}

\begin{figure}
\centering
\includegraphics[scale=0.6]{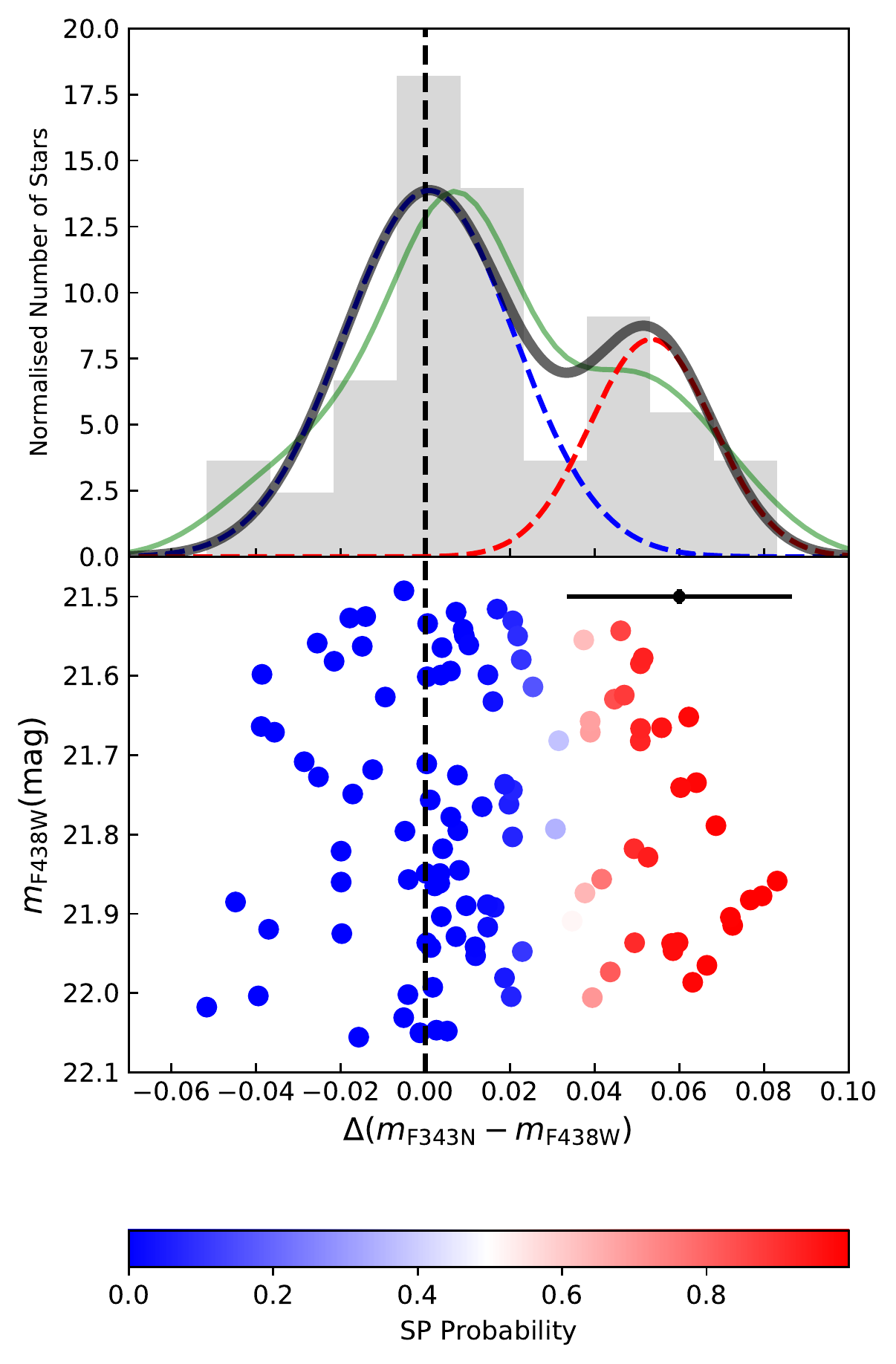}
\caption{{\it Top Panel:} Histogram of the distribution of selected SGB stars in NGC 1978,  in $\Delta(Un-B)$ colours. The black solid line represents the two-component GMM best-fit function to the unbinned data. 
The blue (red) dashed curve represents the first (second) Gaussian component in the fit. The green curve indicates the Kernel Density Estimator (KDE). {\it Bottom Panel:} $\Delta(Un-B)$ vs. $B$ is shown, where stars are colour coded by the probability to belong to the SP. The black dashed vertical line marks the adopted fiducial line. The black errorbar shown in the bottom panel represents the typical error in $\Delta(Un-B)$ colours and $B$ magnitudes. The error on the $B$ magnitude is smaller than the black filled circle.}
\label{fig:pops}
\end{figure}

\section{Constraining the Age difference between the FP/SP Populations on the SGB}
\label{sec:sgb}

In this section we estimate the age difference between the two populations present in NGC 1978 by analysing the SGB.
In \S \ref{subsec:sgbsel} we will select SGB stars and we will assign each star a probability to belong either to a first population or second population, taking advantage of UV CMDs. We will also show that 
this separation is expected from stellar isochrones, if chemical abundance variations are present. 

In \S \ref{subsec:sgban} we will outline the analysis of the SGB and report our results in terms of age difference between the first and second population of stars in the SGB of NGC 1978.

\begin{figure*}
\centering
\includegraphics[scale=0.55]{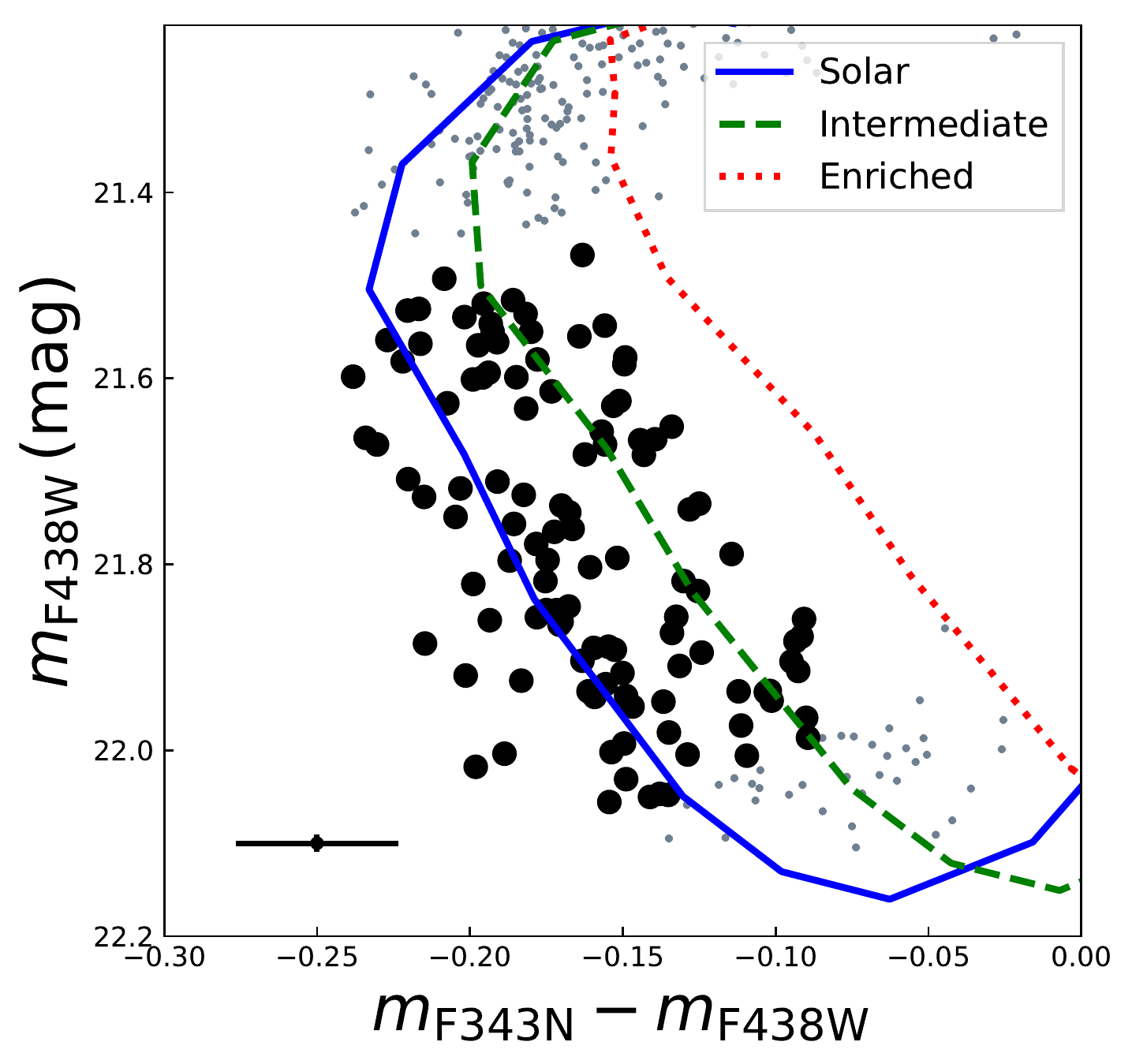}
\includegraphics[scale=0.55]{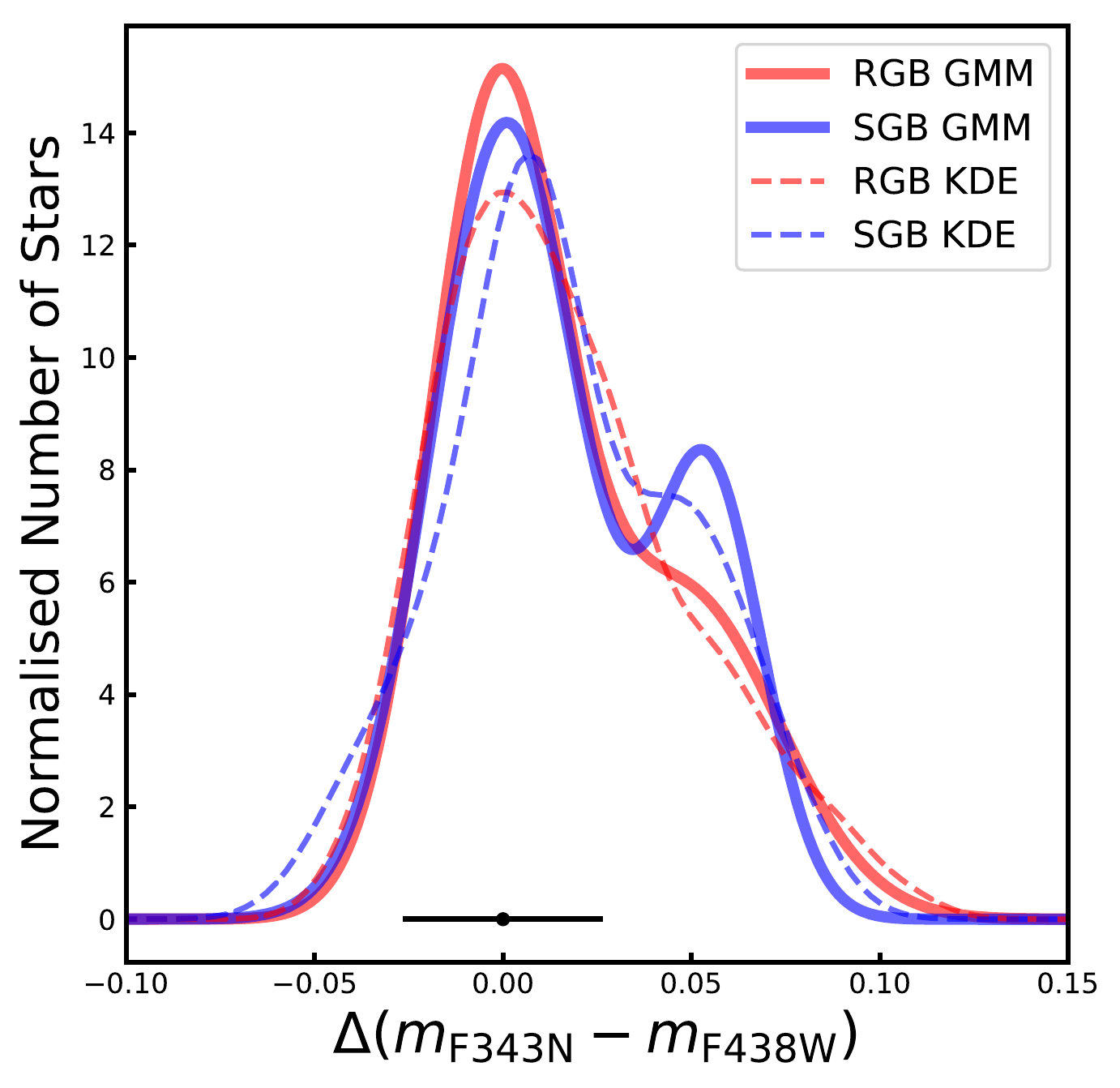}
\caption{\emph{Left panel}: SGB stars in the $Un-B$ vs. $B$ CMD for NGC 1978. The final SGB selected stars are marked with black filled circles. The solid blue, dashed green and dotted red curves represent our stellar atmosphere models from the 2.2 Gyr, [Fe/H]=-0.5 dex MIST isochrone by using solar ([C/Fe]$=$[O/Fe]$=$[N/Fe]=0 dex), intermediate ([C/Fe]$=$[O/Fe]$= -0.1$ dex, [N/Fe]$=+0.5$ dex) and enriched ([C/Fe]$=$[O/Fe]$= -0.6$ dex, [N/Fe]$=+1.0$ dex) chemical abundance mixtures, respectively. \emph{Right panel}: Two-component GMM best-fit function to the unbinned $\Delta(Un-B)$ colours. The red curve is calculated for the selected RGB stars \citep{martocchia17b}, while the blue curve represents the fit to the SGB stars which were selected in this work (see Fig. \ref{fig:pops}). A systematic offset has been applied to the $\Delta(Un-B)$ colours of the RGB such that the GMM's main peak coincided with the SGB GMM. The black errorbar shown in both panels represents the typical error in $\Delta(Un-B)$ colours and $B$ magnitudes. The error on the $B$ magnitude is smaller than the black filled circle The red (blue) dashed line represents the KDE for the RGB (SGB) selected stars.}.
\label{fig:modelsgb}
\end{figure*}

\subsection{SGB stars selection}
\label{subsec:sgbsel}

Firstly, we selected SGB stars in the $V-I$ vs. $I$ CMD of NGC 1978, as the SGB and MS overlap in UV filters. Fig. \ref{fig:sgbsel} shows the $V-I$ vs. $I$ CMD of NGC 1978, with black filled circles marking the first selection of SGB stars. 

We plotted the first selection of SGB stars in the $Un-B$ vs. $B$ CMD (left panel of Fig. \ref{fig:UnB}). We then made the final selection along the relatively vertical part of the SGB, where there is a visible split in the observed sequence. Red filled circles mark the final selected SGB stars in the left panel of Fig. \ref{fig:UnB}. 
Such a selection was made since we expect a split in this region of the SGB, as we will show later on in this Section (see Fig. \ref{fig:modelsgb}).

According to the isochrone describing the shape of the SGB in UV filters (Fig. \ref{fig:modelsgb}), we defined a fiducial line on the blue part of the selected SGB stars in the $Un-B$ vs $B$ CMD and this is displayed in the right panel of Fig. \ref{fig:UnB} as a solid red line. Black filled circles represent the final selected SGB stars in this figure. Next, we calculated the distance in $Un-B$ colours of each SGB star from the fiducial line, $\Delta(Un-B)$. 
Next, we fit the unbinned verticalised $\Delta(Un-B)$ data with two-component Gaussian Mixture Models (GMMs) to identify the presence of multiple Gaussian components in the colour distribution. 
We fit the data with the SCIKIT-LEARN python package called MIXTURE\footnote{\url{http://scikit-learn.org/stable/modules/mixture.html}}, which applies the expectation-maximization algorithm for fitting mixture-of-Gaussian models. The result of the fit is shown as a solid black line in the top panel of Fig. \ref{fig:pops} over the histogram of the distribution in $\Delta(Un-B)$ colours. For comparison, we also show the non-parametric Kernel Density Estimate (KDE) to the unbinned data as a green curve. The adaptive bandwidth was calculated by using the Scott's rule of thumb \citep{scott}.
The blue (red) dashed curve represents the first (second) Gaussian component in the fit. We will refer to the blue component as representative of a first population (FP) in the cluster, and to the red component as representative of a second population (SP). 

For each star, we assigned a probability to belong to the FP and to the SP by using the respective Gaussian function found by the GMM fit. 
The bottom panel of Fig. \ref{fig:pops} shows the $\Delta(Un-B)$ colours vs. $B$, where the stars are colour coded by the probability to belong
to the SP. The black dashed vertical line marks the adopted fiducial line.

The left panel of Fig. \ref{fig:modelsgb} shows the $Un-B$ vs. $B$ CMD of the SGB stars in NGC 1978. 
Final SGB selected stars are indicated with black filled circles.
We compare our data with three isochrones superimposed. We extensively described how our stellar N-enriched isochrones are built in \cite{paperIII,martocchia17b} and refer the interested reader to them for future details. The blue solid, green dashed, red dotted curves represent theoretical isochrones, derived by using the 2.2 Gyr, [Fe/H]=-0.5 dex MIST isochrone \citep{dotter16,choi16}, for three different chemical abundance mixtures, denominated as solar ([C/Fe]$=$[O/Fe]$=$[N/Fe]=0 dex), intermediate ([C/Fe]$=$[O/Fe]$= -0.1$ dex, [N/Fe]$=+0.5$ dex) and enriched ([C/Fe]$=$[O/Fe]$= -0.6$ dex, [N/Fe]$=+1.0$ dex) models, respectively. For the N-enhanced isochrones, the choice of C and O abundances were chosen to keep the [(C+N+O)/Fe] constant between the models, according to what is observed in GCs (e.g. \citealt{carretta05}).

The observed split in the SGB of the $Un-B$ vs. $B$ CMD is consistent with expectations from theoretical isochrones. 
 Fig. \ref{fig:modelsgb} reveals that a clear split is expected in $Un-B$ colours if chemical variations are present in the SGB, which is consistent to what we observe in the data. Note that for $B\lesssim 21.5$ the split in the SGB is less evident and the isochrones tend to merge for smaller magnitudes, while for $B\gtrsim 22.1$ models with different abundance mixtures intersect. This explains the adopted selection which includes only the relatively vertical part of the SGB. Additionally, based on the RGB analysis in \cite{martocchia17b}, NGC 1978 is expected to have an intermediate N-enrichment ([N/Fe] = $ +0.5$ dex), which is the same enrichment we obtain when comparing data to the isochrones with different chemical mixtures in the SGB (Fig. \ref{fig:modelsgb}). 
 
The right panel of Fig. \ref{fig:modelsgb} shows the two-component GMM best-fit function to the unbinned $\Delta(Un-B)$ colours for both RGB (red curve) and SGB (blue curve) stars. Also, the comparison among the KDEs is shown. It is clear from this figure that both SGB and RGB show the same broadness in $\Delta(Un-B)$ colours and very similar GMM best-fit distributions and KDEs.

\subsection{SGB Analysis}
\label{subsec:sgban}

\begin{figure*}
\centering
\includegraphics[scale=0.37]{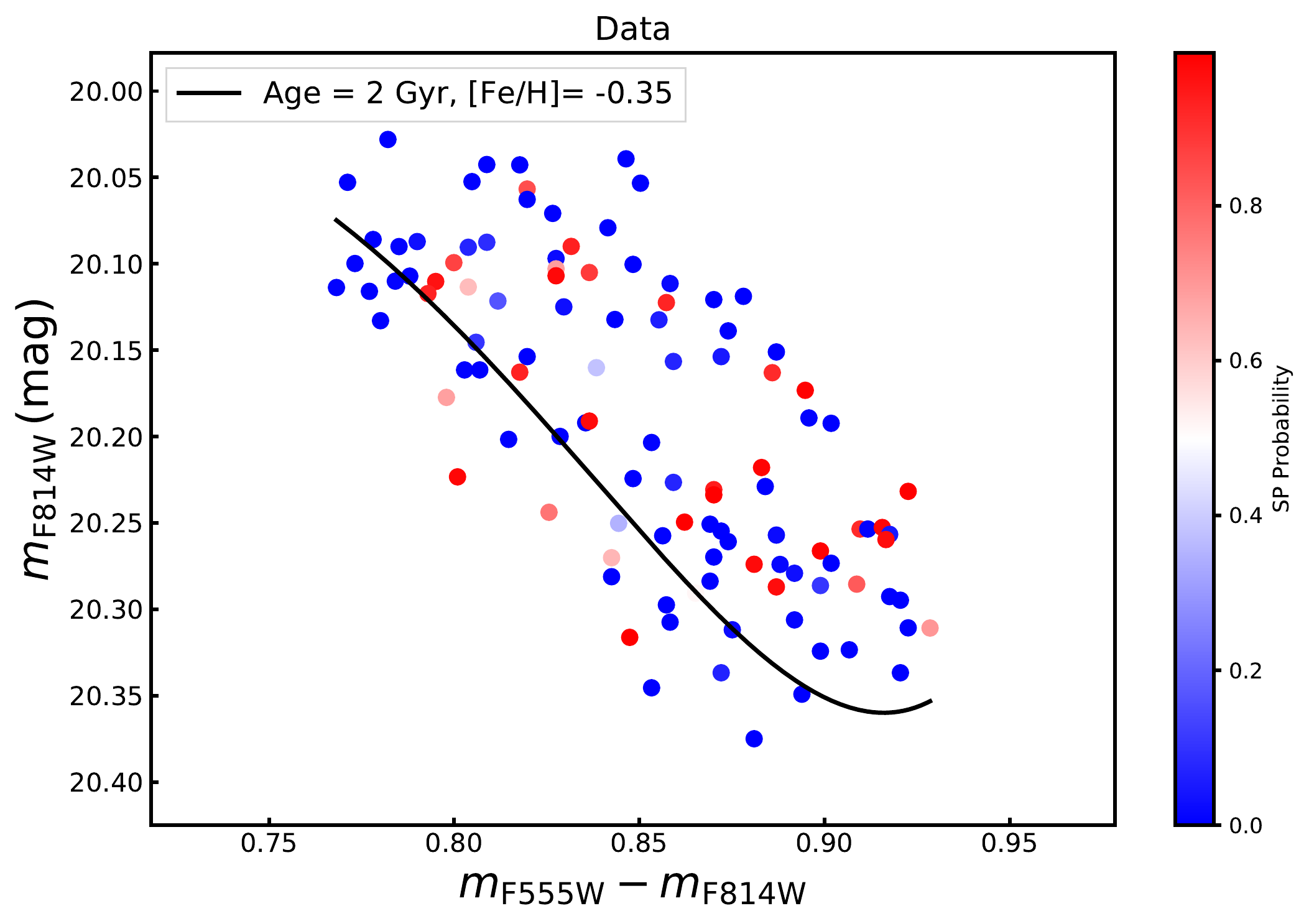}
\includegraphics[scale=0.37]{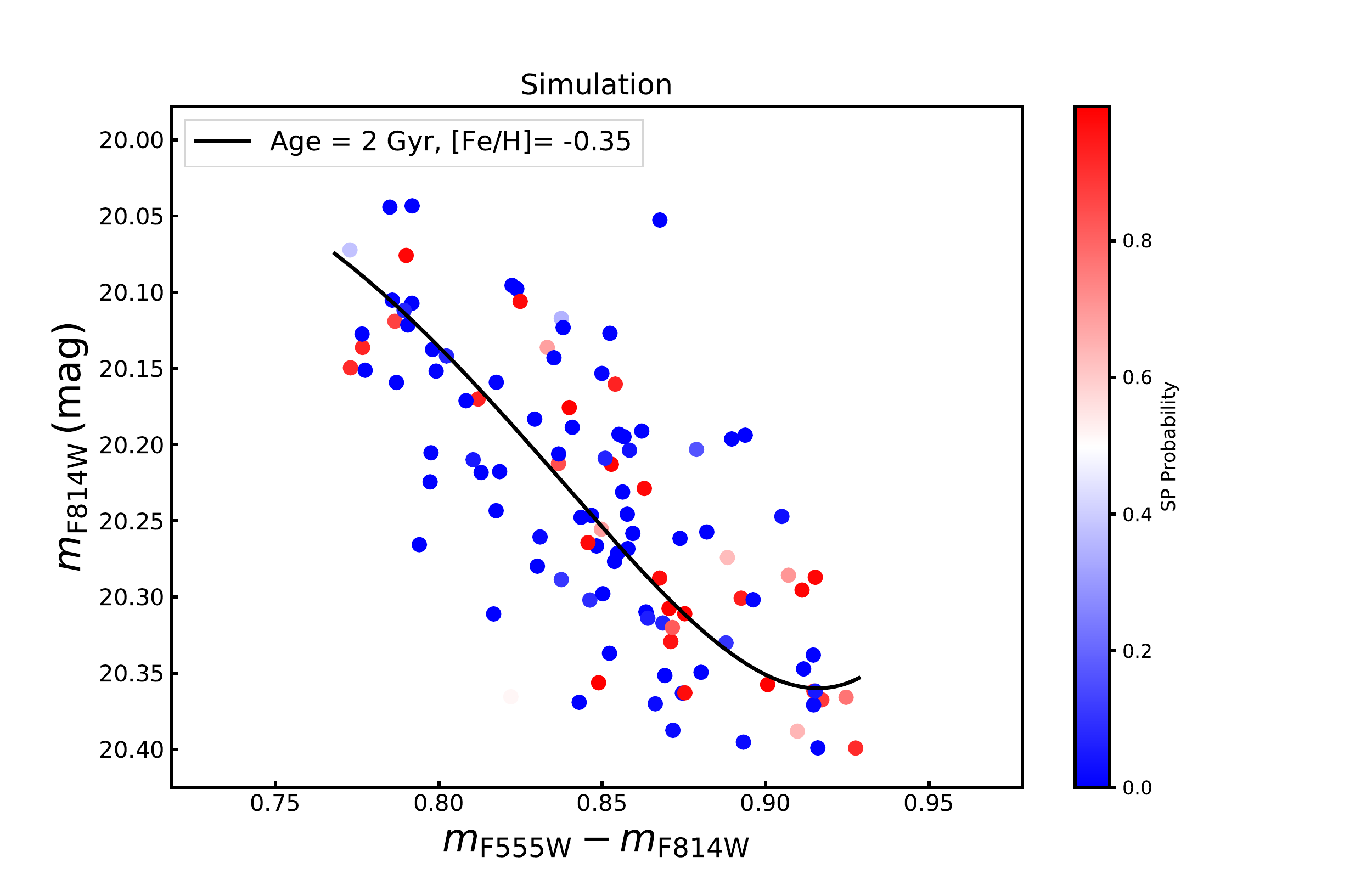}
\caption{{\it Left panel:} SGB selected stars in the $V-I$ vs. $I$ CMD for NGC 1978 (left panel) and MC simulation of our SGB data, where observational errors are taken into account (right panel). Stars are colour-coded according to the probability of belonging to the SP. The black solid line indicates the defined fiducial line on the 2 Gyr, [Fe/H]$= -0.35$ dex BaSTI isochrone on the SGB (Fig. \ref{fig:iso}).}
\label{fig:datasimu}
\end{figure*}

As a first step, we plotted the SGB selected stars (see \S \ref{subsec:sgbsel}) on the $V-I$ vs. $I$ CMD. This is shown in the left panel of Fig. \ref{fig:datasimu}, where stars are colour-coded by the probability to belong to the SP. In this filter combination, no spreads or split are expected due to the presence of MPs (e.g., \citealt{sbordone11}). In optical filters, the SGB is expected not to be sensitive to MPs assuming that the CNO sum is constant, however the magnitude of the SGB is a strong function of age, hence if there is an age difference between the FP and SP, their SGB magnitudes should differ (e.g., \citealt{li14}). Star-to-star variations of He abundance might generate an offset on the SGB as well. However, significant He spreads in standard GCs are usually associated with a large N enrichment (e.g., \citealt{milone15}, \citealt{milone17}). Assuming that young star clusters behave as old GCs, no remarkable He spreads are to be expected in NGC 1978, as the N enhancement in NGC 1978 appears to be small \citep{martocchia17b}.

A first look at the left panel of Fig. \ref{fig:datasimu} reveals that no clear and visible offset is observed between the two populations. 
In order to better quantify the presence of a possible offset, we defined a fiducial line on the 2 Gyr, [Fe/H]$= -0.35$ dex BaSTI isochrone on the SGB (Fig. \ref{fig:iso}). This is shown as a black solid curve in both panels of Fig. \ref{fig:datasimu}.
We calculated the distance in $I$ magnitudes of each SGB star from the fiducial line, $\Delta I$. 
Then, we calculated the weighted mean in $\Delta I$ for the FP and SP by using the probability to belong to the FP and SP as weights.
Our observed age difference between the two populations, in units of magnitude, results to be $\Delta$Mag$_{OBS} \equiv < \Delta I_{\rm FP}> - <\Delta I_{\rm SP}> = 0.0027$ mag for the $V-I$ vs. $I$ CMD.

\begin{figure*}
\centering
\includegraphics[scale=0.45]{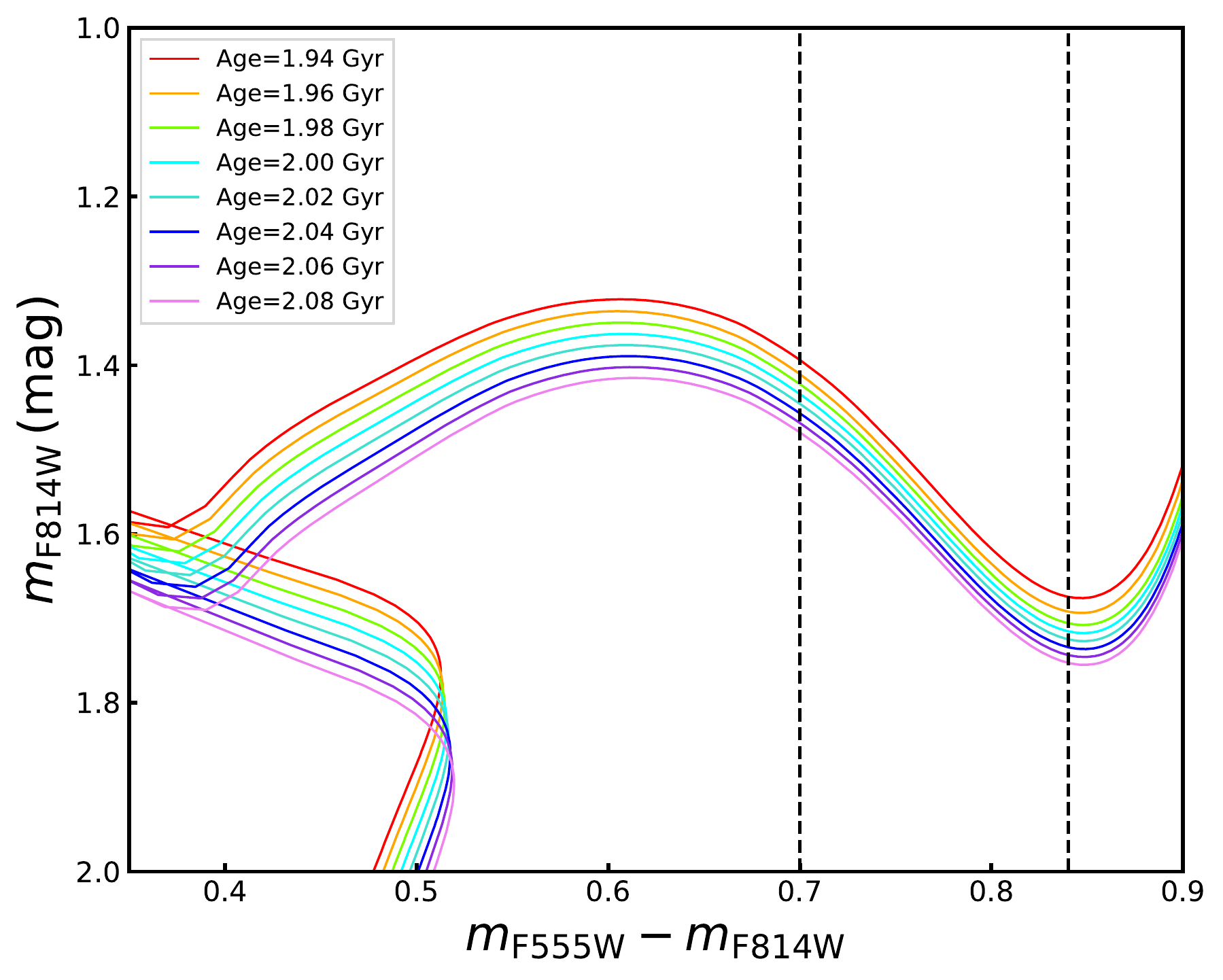}
\includegraphics[scale=0.45]{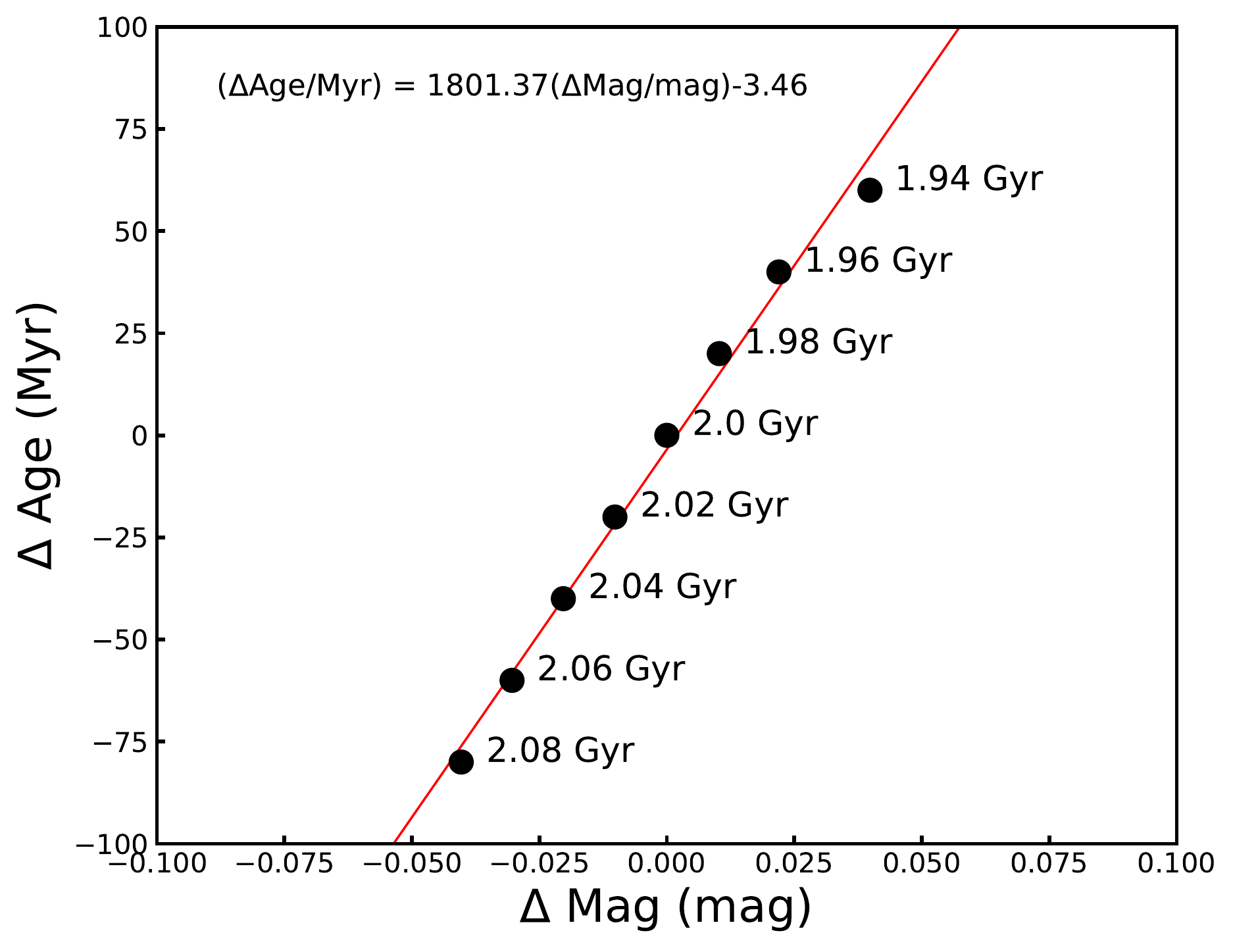}
\caption{{\it Left Panel:} BaSTI isochrones in the $V-I$ vs. $I$ CMD with ages ranging from 1.94 up to 2.08 Gyr, spaced by 20 Myr ([Fe/H]$=-0.35$ dex). The vertical black dashed lines highlight the SGB region used to calculate the expected and observed age differences between the two populations. {\it Right Panel:} $\Delta$Mag vs. $\Delta$Age relation (red solid line) for the SGB when the isochrones in the left panel are taken into account. See text for more details.}
\label{fig:isocomp}
\end{figure*}

Hence, we can compare the observed age difference in magnitude to isochrones in the SGB to obtain the age difference in units of Myr.
As discussed in \S \ref{sec:observations}, we used BaSTI isochrones. The left panel of Fig. \ref{fig:isocomp} shows the isochrones we used for the calculation, from age = 1.94 Gyr up to 2.08 Gyr, spaced by 20 Myr. 
Two vertical dashed lines mark the selected region of the SGB where we performed the calculation, corresponding to the observed section of the SGB where the split is seen in UV colours. Firstly, we used the 2 Gyr isochrone as an age reference for NGC 1978 (see Fig. \ref{fig:iso}). 
We then calculated the mean difference in $I$ magnitudes between the 2 Gyr isochrone and the other isochrones (displayed in Fig. \ref{fig:isocomp}), in the SGB. In this way, 
we can transform the observed difference in magnitude to age differences (defined as $\Delta$Age).

These values are reported in the right panel of Fig. \ref{fig:isocomp}. Next, we fit the values with a linear relation and we found the following:

\begin{equation}
(\Delta {\rm Age/Myr}) =  1801.37 \times (\Delta {\rm Mag/mag}) -3.46.
\label{eq:deltaagemag}
\end{equation}

The best-fit is shown as a red solid line in the right panel of Fig. \ref{fig:isocomp}. 
We then used this relation to convert our observed ($\Delta$Mag$_{OBS}$) value into age differences in Myr between the two populations.

We found that the age difference in the $V-I$ vs. $I$ CMD between the FP and SP is 1.4 Myr.

We used Monte Carlo (MC) simulations to estimate the uncertainty of this result.
Ideally, we would sample stars from two separate isochrones for the FP and SP and calculate the inferred age difference with respect to the input age difference assuming that the magnitude spread of each population is solely due to the photometry uncertainties.
However, the measured value of 1.4\,Myr is much smaller than the age resolution of the isochrone grid (20\,Myr) and by all means consistent with null intrinsic spread. In this situation we need to investigate the level of stochasticity introduced in this result due to the limited number of available stars. Thus, we sampled from a single isochrone model the same number of stars as the ones used in the fit, using their photometric errors, assigned to them the same probability distribution of belonging to the FP and the SP as in the real data, and repeated this process 100,000 times. 
The right panel of Fig. \ref{fig:datasimu} shows an example of a simulation, which appears to well reproduce our observations. 
According to Fig. \ref{fig:photerr}, we considered as photometric errors for the $V$ and $I$ filters, the values of $\sigma_V$ and $\sigma_I$ found for the inner centre of the cluster ($r<10''$), which also correspond to the maximum values as a function of distance.
We found that the Monte Carlo distribution has a mean of 0\,mag, as expected, and $\sigma=0.013$\,mag, that corresponds to an age difference of 19.9\,Myr (by using Eq. \ref{eq:deltaagemag}).

We found that the age difference between the two populations in the SGB is then $1\pm20$ Myr, which is consistent with zero. 
If different stellar isochrones are used, we obtain similar constraints. When the MIST models are used (however with a resolution of 40 Myr), we find a measured difference of $5\pm29$ Myr. 
This is an extremely tight constraint for the origin of MPs and it shows that the two populations have essentially the same age. We will discuss this in detail in \S \ref{sec:discussion}.

We repeated the analysis described above, using the $B-I$ vs. $I$ CMD. In this case, the observed age difference between the two populations, in units of magnitude, is 
$\Delta$Mag$_{OBS} = 0.0017$ mag. From the simulations, the expected magnitude difference between the two populations is a Gaussian with peak at zero and a $\sigma = 0.01$ mag. 
By comparison with $B-I$ vs. $I$ BaSTI isochrones, we found that the age difference between FP and SP is consistent with zero, being $0.5\pm14.7$ Myr when an age of 2 Gyr is considered as a reference for the calculation.

One may argue that absolute cluster ages are not well-known due to the degeneracy established by the numerous parameters involving the isochrone fitting. 
In this case, as a further test for our results, we adopted two other ages as reference age. We did the same isochrone comparison by using an age of 2.5 Gyr and 1.8 Gyr. 
We then calculated the linear relations among $\Delta$Mag and $\Delta$Age and computed the age difference in Myr between FP and SP by using these new best-fits.
We still found that the age difference is consistent with the results from our original experiment.

\begin{figure}
\centering
\includegraphics[scale=0.45]{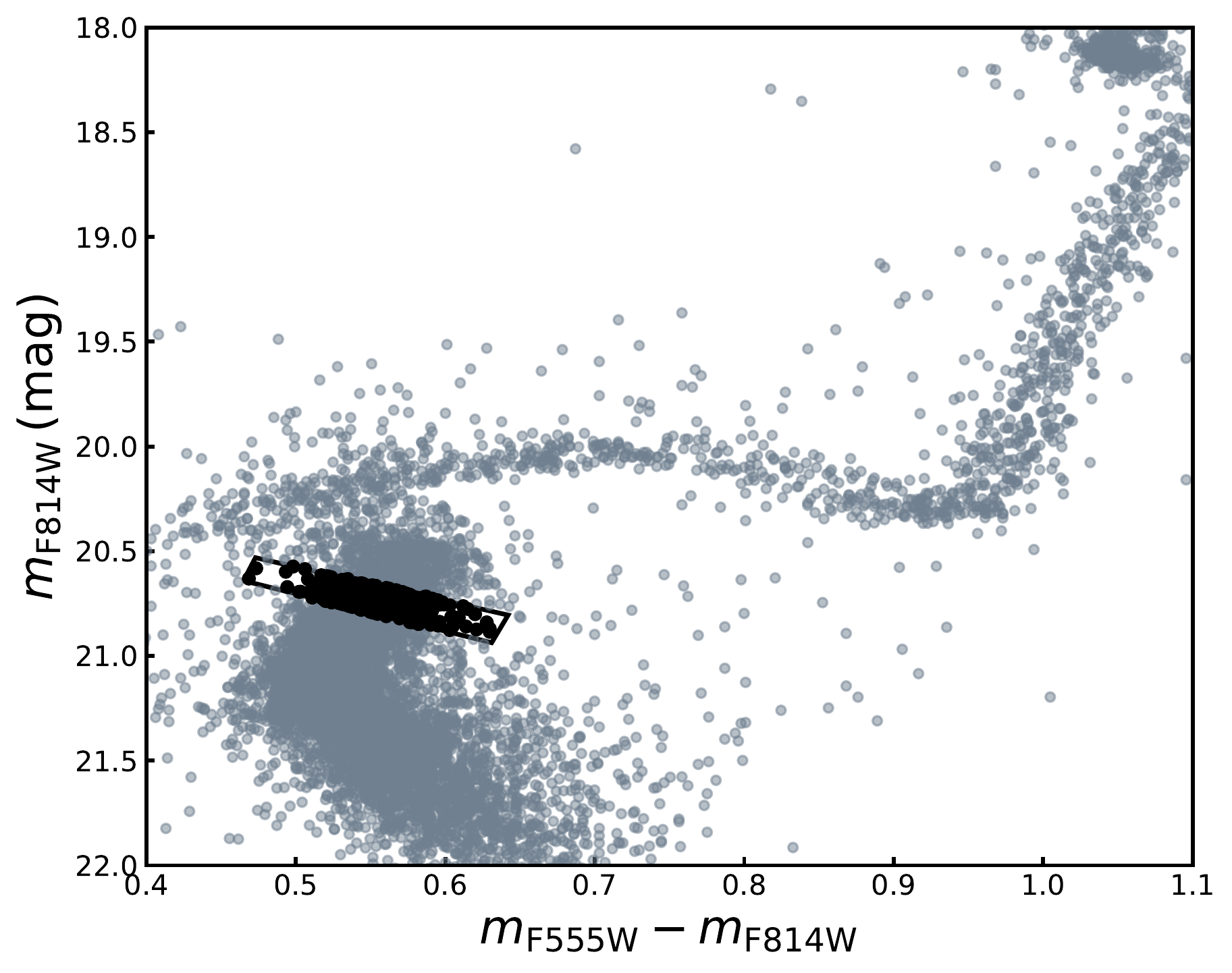}
\caption{$V-I$ vs. $I$ CMDs of NGC 1978. The black box indicates the locus of the selection of MS stars, marked by black filled circles.} 
\label{fig:mssel}
\end{figure}

\section{The Main Sequence Turnoff}
\label{sec:sfh}

In this Section, we will outline the analysis of the MSTO width of NGC 1978.

\begin{figure*}
\centering
\includegraphics[scale=0.55]{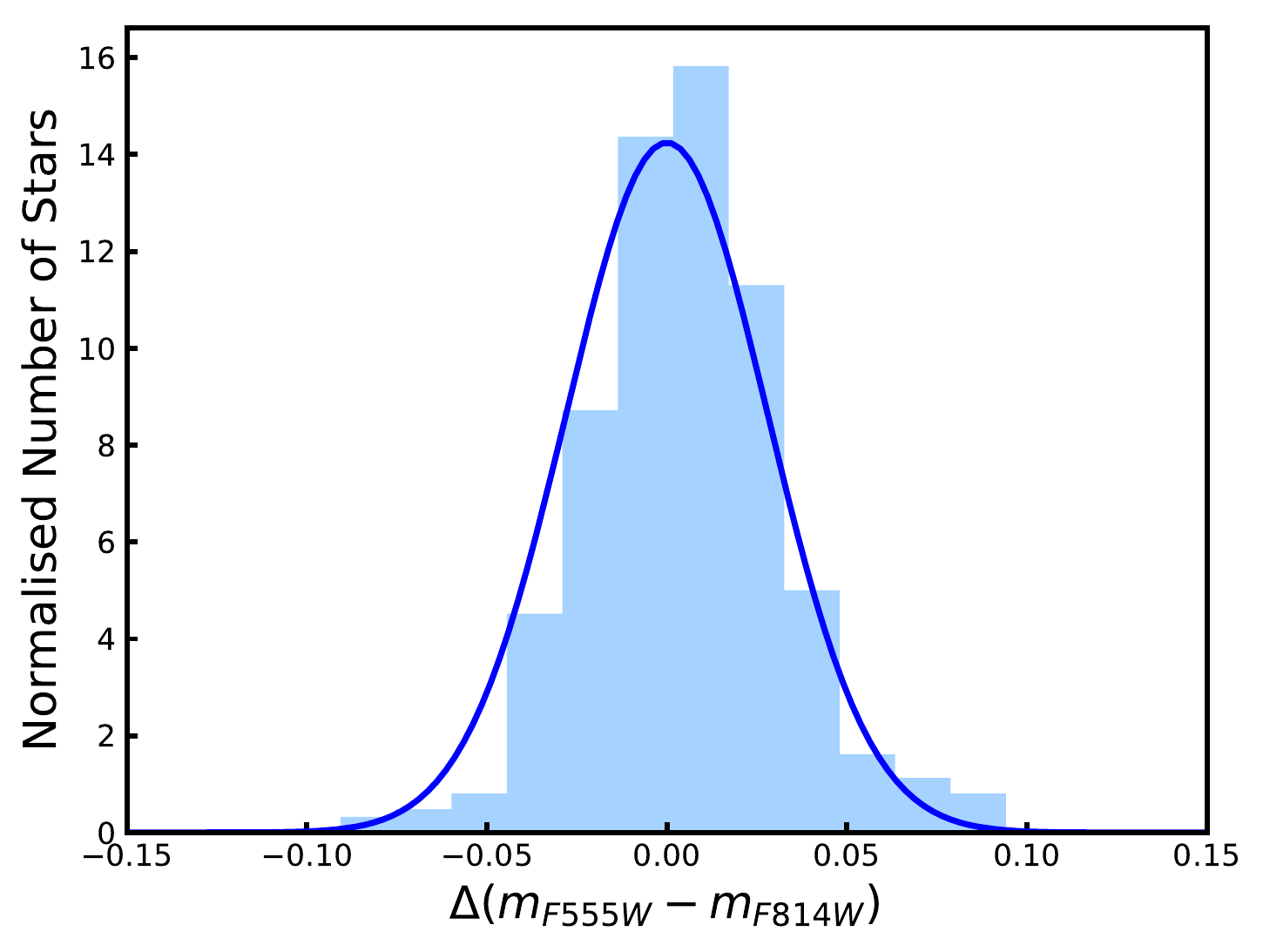}
\includegraphics[scale=0.55]{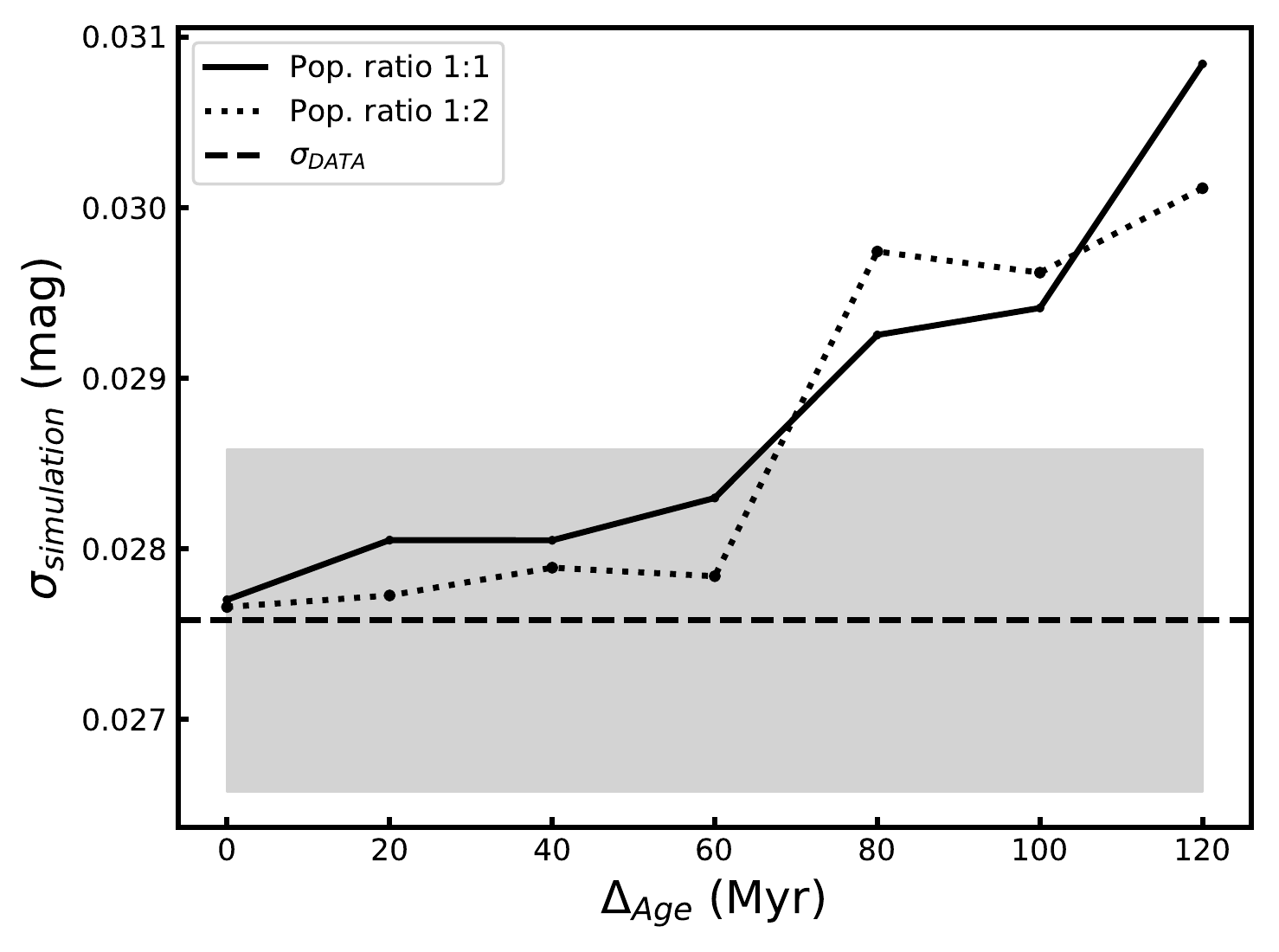}
\caption{{\it Left panel:} Histogram of the distribution of the selected MS stars in $\Delta(V-I)$ colours. The blue solid line indicates a Gaussian function centred at zero and with $\sigma =$ 0.028, which represents the observational error obtained from the AS test. {\it Right panel:} $\Delta_{Age}$ as a function of $\sigma_{simulation}$ for a population ratio of 1:1 (black solid line) and 1:2 (black dotted line). The black dashed line indicates the value of $\sigma_{DATA}$, while the gray shaded area marks the uncertainty on $\sigma_{DATA}$. See text for more details.}
\label{fig:simMS}
\end{figure*}

We selected MS stars in the $V-I$ vs. $I$ CMD. Fig. \ref{fig:mssel} shows the $V-I$ vs. $I$ CMD of NGC 1978 with black filled circles indicating the MS selected stars. 
Next, we defined a fiducial line in the $V-I$ vs. $I$ space. We calculated the distance in $V-I$ colours of each star from the fiducial line, $\Delta (V-I)$. The left panel of Fig. \ref{fig:simMS} displays the histogram of the distribution of the selected MS stars in $\Delta(V-I)$ colours.
The blue solid line indicates a Gaussian function centred at zero and with $\sigma =$ 0.028, which represents the error in $\Delta(V-I)$ colours obtained from the AS experiment by applying the same selection box that was used for the data (Fig. \ref{fig:mssel}).  This comparison immediately shows that the observed spread in the MS $V-I$ colours can be entirely attributed to observational errors.

Next, we used mock samples from theoretical isochrones in order to compare the MSTO width of the data with simulations. Thus, we can derive an upper limit on the age spread hidden in the observational errors. We randomly sampled 10,000 stars from the 2 Gyr BaSTI isochrone (Fig. \ref{fig:iso}) in the MSTO region\footnote{Random stars were sampled in the colour-magnitude space 0.42$\lesssim V-I \lesssim$ 0.62, 20.4 $\lesssim I \lesssim$ 21.1.}, by adding observational errors both in magnitude and colours. We then applied to the simulation the same selection cut applied to the data (Fig. \ref{fig:mssel}). For this set of data, we estimated the distance in $V-I$ colours of each star from the isochrone, $\Delta (V-I)$; finally, we calculated the standard deviations of stars in $\Delta (V-I)$ colours, namely $\sigma_{simulation}$.
We then did this analysis by sampling 10,000 random stars from the 2 Gyr BaSTI isochrone and 10,000 star from another isochrone simultaneously, imitating two populations separated in age by $\Delta(age)$. We repeated this with isochrones spaced by 20, 40, 60, 80, 100 and 120 Myr. Each simulation was treated in exactly the same way as the observations. We calculated $\sigma_{simulation}$ for each sample and we show $\Delta_{Age}$ as a function of $\sigma_{simulation}$ in the right panel of Fig. \ref{fig:simMS}. 
We calculated our $\sigma_{DATA}$ from the $\Delta(V-I)$ colours for a direct comparison. This value is shown as a black dashed line in the right panel of Fig. \ref{fig:simMS}. 
The gray shaded area marks the uncertainty on $\sigma_{DATA}$, which was calculated on the unbinned data by using a bootstrap technique \citep{efron93} based on 10,000 realizations.

By making a comparison between $\sigma_{DATA}$ and $\sigma_{simulation}$ (right panel of Fig. \ref{fig:simMS}), we can put an upper limit of $\sim$65 Myr to an age spread on the MSTO of NGC 1978, at 2$\sigma$ confidence level. We discuss this result in detail in \S \ref{sec:discussion}.

Additionally, based on the results by \cite{goudfrooij11}, we repeated the same simulations when considering a population ratio of 1:2, i.e. by sampling 10,000 random stars for the younger population and 5,000 stars for the older population. This is shown as a black dotted line in the right panel of Fig. \ref{fig:simMS} and it is consistent with what we found
with a 1:1 population ratio. However, we note that this scenario would be inconsistent with the AGB scenario due to the mass-budget problem (e.g., \citealt{bastianlardo2018}).

\section{Discussion and Conclusions}
\label{sec:discussion}

The aim of this work was to exploit the unique characteristics of NGC 1978 to place stringent constraints on the presence of a spread in age amongst its multiple constituent stellar populations. 

We took advantage of the structure of its SGB to estimate the age difference between the two subpopulations present in NGC 1978. 
We find an age difference between the two populations of $1\pm20$ Myr, when the $V-I$ vs. $I$ CMD is taken into account. If we repeat the same analysis by taking into account the $B-I$ vs. $I$ CMD we find an age difference of $0.5\pm14.7$ Myr (\S \ref{subsec:sgban}).
   
From this analysis, it emerges that the two populations present in NGC 1978 have the same age, or at most their age difference is very little. 

Such results establish very tight constraints on the onset of multiple populations and provide limits on the nature of the polluters.

Models for the origin of multiple populations which adopt multiple generations of star-formation using AGB stars as the source of polluting material, predict an age difference of at least 30 Myr between the 1st and 2nd populations, potentially being as large as 200 Myr.
Our results show instead that the FP and SP in NGC 1978 formed at the same time or very close to each other, creating significant tension with predictions from this family of models. 

Currently all models put forward for the nature of multiple populations present serious drawbacks in reproducing the very complicated details provided by the observations (\citealt{bastian15}, \citealt{renzini15}, \citealt{prantzos17}, \citealt{bastianlardo2018}). 
However, the results presented here support a scenario where no multiple bursts of star formation are invoked or that they happened nearly concurrently, i.e. abundance anomalies are not originated by means of multiple generations created over a large time separation.

Additionally, we estimated the broadness of the MSTO of NGC 1978. While most LMC and SMC clusters studied to date, aged less than 2 Gyr, clearly show an extended MSTO (e.g. \citealt{mackey08,milone09}), we find that NGC 1978 does not show a prominent eMSTO, i.e. the observed spread in the TO is comparable with photometric errors (\S \ref{sec:sfh}). We can put an upper limit of $\sim$ 65 Myr to the age spread in the MSTO.

The origin of the eMSTO in young star clusters is strongly debated in the community. However, NGC 1978 is a peculiar and interesting case which might lead to a major breakthrough. The nature of the eMSTO  was originally explained by the presence of age spreads of the order of 200-800 Myr within a cluster. Such age spreads, perhaps related to the cluster's high escape velocity \citep{goudfrooij14}, require the occurrence of multiple bursts of star formation over an extended period within the cluster.

\cite{bastiandemink09} proposed that alternatively the eMSTO may be caused by a range of stellar rotation rates in a single-aged population.
Predictions and comparisons with observations for the stellar rotation scenario were made by studying eMSTO populations of both intermediate age \citep{brandt15} and young massive clusters (YMCs, \citealt{niederhofer15}). The latter found the so-called $\Delta$(Age) vs. Age relation, i.e. the expected/inferred age spread within a cluster is directly proportional to the age of the cluster.
This trend was also corroborated at younger ages \citep{milone15,bastian16} and it is in good agreement with predictions of the stellar rotation scenario.
Such a relation and the results presented here are clearly at odds with predictions of the age spread scenario as the origin of the eMSTO. 

We find that chemical abundance variations are present in the RGB of NGC 1978, whereas no MPs are detected in YMCs or intermediate age clusters that show prominent eMSTOs($< 2$ Gyr, e.g. \citealt{mucciarelli14,cabrera16b,paperIII, martocchia17b}). Among these we can find the SMC cluster NGC 419, which has very similar properties to NGC 1978 (mass, radius), but it is $\sim$ 500 Myr younger and show one of the largest eMSTO. 

On the contrary, we do not expect the presence of an eMSTO in NGC 1978 according to the stellar rotation scenario. \cite{brandt15} explicitly predicted a turnover in the distribution, so that by an age of $\sim$ 2 Gyr the eMSTO should disappear (in the rotation scenario) which agrees with the results presented here. Stellar masses for TO stars at this age ($\sim 2$ Gyr) drop below 1.5 \msun, such that stars are magnetically braked and become slow rotators. This is fully consistent with the observations, thus the work reported here support a scenario where the eMSTO is caused by a stellar rotation effect.

\section*{Acknowledgments}

We, in particular, N.B., I.P. and V. K.-P., gratefully acknowledge financial support for this project provided
by NASA through grant HST-GO-14069 for the Space Telescope Science Institute, which is operated by the Association
of Universities for Research in Astronomy, Inc., under NASA contract NAS526555. 
F.N. acknowledges support from the European Research Council (ERC) under European Union's Horizon 2020 research and innovation programme (grant agreement No 682115).
C.L. thanks the Swiss National Science Foundation for supporting this research through the Ambizione grant number PZ00P2\_168065. N.B. gratefully acknowledges financial support from the Royal Society (University Research Fellowship) and the European Research Council
(ERC-CoG-646928-Multi-Pop). 
C.U. gratefully acknowledges financial support from European Research Council (ERC-CoG-646928-Multi-Pop).
D.G. gratefully acknowledges support from the Chilean BASAL Centro de Excelencia en Astrof\'isica
y Tecnolog\'ias Afines (CATA) grant PFB-06/2007. 
Support for this work was provided by NASA through Hubble Fellowship grant \# HST-HF2-51387.001-A awarded by the Space Telescope Science Institute, which is operated by the Association of Universities for Research in Astronomy, Inc., for NASA, under contract NAS5-26555.

\bibliographystyle{mn2e}
\bibliography{sfh.bib}

\begin{thebibliography}{48}
\expandafter\ifx\csname natexlab\endcsname\relax\def\natexlab#1{#1}\fi

\bibitem[{{Bastian} {et~al}\mbox{.}(2013){Bastian}, {Cabrera-Ziri}, {Davies},
  \& {Larsen}}]{bastian13}
{Bastian} N., {Cabrera-Ziri} I., {Davies} B., {Larsen} S.~S., 2013, MNRAS, 436,
  2852

\bibitem[{{Bastian}, {Cabrera-Ziri} \& {Salaris}(2015){Bastian},
  {Cabrera-Ziri}, \& {Salaris}}]{bastian15}
{Bastian} N., {Cabrera-Ziri} I., {Salaris} M., 2015, MNRAS, 449, 3333

\bibitem[{{Bastian} \& {de Mink}(2009)}]{bastiandemink09}
{Bastian} N., {de Mink} S.~E., 2009, MNRAS, 398, L11

\bibitem[{{Bastian} \& {Lardo}(2017)}]{bastianlardo2018}
{Bastian} N., {Lardo} C., 2017, ArXiv e-prints, 1712.01286

\bibitem[{{Bastian} {et~al}\mbox{.}(2016){Bastian}, {Niederhofer},
  {Kozhurina-Platais}, {Salaris}, {Larsen}, {Cabrera-Ziri}, {Cordero},
  {Ekstr{\"o}m}, {Geisler}, {Georgy}, {Hilker}, {Kacharov}, {Li}, {Mackey},
  {Mucciarelli}, \& {Platais}}]{bastian16}
{Bastian} N. {et~al.}, 2016, MNRAS, 460, L20

\bibitem[{{Bellazzini} {et~al}\mbox{.}(2002){Bellazzini}, {Fusi Pecci},
  {Montegriffo}, {Messineo}, {Monaco}, \& {Rood}}]{bellazzini02}
{Bellazzini} M., {Fusi Pecci} F., {Montegriffo} P., {Messineo} M., {Monaco} L.,
  {Rood} R.~T., 2002, \aj, 123, 2541

\bibitem[{{Brandt} \& {Huang}(2015)}]{brandt15}
{Brandt} T.~D., {Huang} C.~X., 2015, ApJ, 807, 25

\bibitem[{{Cabrera-Ziri} {et~al}\mbox{.}(2016{\natexlab{a}}){Cabrera-Ziri},
  {Bastian}, {Hilker}, {Davies}, {Schweizer}, {Kruijssen},
  {Mej{\'{\i}}a-Narv{\'a}ez}, {Niederhofer}, {Brandt}, {Rejkuba}, {Bruzual}, \&
  {Magris}}]{cabrera16}
{Cabrera-Ziri} I. {et~al.}, 2016{\natexlab{a}}, \mnras, 457, 809

\bibitem[{{Cabrera-Ziri} {et~al}\mbox{.}(2016{\natexlab{b}}){Cabrera-Ziri},
  {Lardo}, {Davies}, {Bastian}, {Beccari}, {Larsen}, \&
  {Hernandez}}]{cabrera16b}
{Cabrera-Ziri} I., {Lardo} C., {Davies} B., {Bastian} N., {Beccari} G.,
  {Larsen} S.~S., {Hernandez} S., 2016{\natexlab{b}}, \mnras, 460, 1869

\bibitem[{{Cannon} {et~al}\mbox{.}(1998){Cannon}, {Croke}, {Bell}, {Hesser}, \&
  {Stathakis}}]{cannon98}
{Cannon} R.~D., {Croke} B.~F.~W., {Bell} R.~A., {Hesser} J.~E., {Stathakis}
  R.~A., 1998, MNRAS, 298, 601

\bibitem[{{Carretta} {et~al}\mbox{.}(2009){Carretta}, {Bragaglia}, {Gratton},
  {Lucatello}, {Catanzaro}, {Leone}, {Bellazzini}, {Claudi}, {D'Orazi},
  {Momany}, {Ortolani}, {Pancino}, {Piotto}, {Recio-Blanco}, \&
  {Sabbi}}]{carretta09}
{Carretta} E. {et~al.}, 2009, Astron. Astrophys., 505, 117

\bibitem[{{Carretta} {et~al}\mbox{.}(2005){Carretta}, {Gratton}, {Lucatello},
  {Bragaglia}, \& {Bonifacio}}]{carretta05}
{Carretta} E., {Gratton} R.~G., {Lucatello} S., {Bragaglia} A., {Bonifacio} P.,
  2005, \aap, 433, 597

\bibitem[{{Choi} {et~al}\mbox{.}(2016){Choi}, {Dotter}, {Conroy}, {Cantiello},
  {Paxton}, \& {Johnson}}]{choi16}
{Choi} J., {Dotter} A., {Conroy} C., {Cantiello} M., {Paxton} B., {Johnson}
  B.~D., 2016, ApJ, 823, 102

\bibitem[{{Conroy} \& {Spergel}(2011)}]{conroyspergel11}
{Conroy} C., {Spergel} D.~N., 2011, \apj, 726, 36

\bibitem[{{Dalessandro} {et~al}\mbox{.}(2015){Dalessandro}, {Ferraro},
  {Massari}, {Lanzoni}, {Miocchi}, \& {Beccari}}]{dalessandro15}
{Dalessandro} E., {Ferraro} F.~R., {Massari} D., {Lanzoni} B., {Miocchi} P.,
  {Beccari} G., 2015, \apj, 810, 40

\bibitem[{{Dalessandro} {et~al}\mbox{.}(2016){Dalessandro}, {Lapenna},
  {Mucciarelli}, {Origlia}, {Ferraro}, \& {Lanzoni}}]{dalessandro16}
{Dalessandro} E., {Lapenna} E., {Mucciarelli} A., {Origlia} L., {Ferraro}
  F.~R., {Lanzoni} B., 2016, ApJ, 829, 77

\bibitem[{{de Mink} {et~al}\mbox{.}(2009){de Mink}, {Pols}, {Langer}, \&
  {Izzard}}]{demink09}
{de Mink} S.~E., {Pols} O.~R., {Langer} N., {Izzard} R.~G., 2009, Astron.
  Astrophys., 507, L1

\bibitem[{{Decressin} {et~al}\mbox{.}(2007){Decressin}, {Meynet}, {Charbonnel},
  {Prantzos}, \& {Ekstr{\"o}m}}]{decressin07}
{Decressin} T., {Meynet} G., {Charbonnel} C., {Prantzos} N., {Ekstr{\"o}m} S.,
  2007, Astron. Astrophys., 464, 1029

\bibitem[{{Denissenkov} \& {Hartwick}(2014)}]{denissenkov14}
{Denissenkov} P.~A., {Hartwick} F.~D.~A., 2014, \mnras, 437, L21

\bibitem[{{D'Ercole} {et~al}\mbox{.}(2008){D'Ercole}, {Vesperini}, {D'Antona},
  {McMillan}, \& {Recchi}}]{dercole08}
{D'Ercole} A., {Vesperini} E., {D'Antona} F., {McMillan} S.~L.~W., {Recchi} S.,
  2008, MNRAS, 391, 825

\bibitem[{{Dotter}(2016)}]{dotter16}
{Dotter} A., 2016, ApJS, 222, 8

\bibitem[{Efron \& Tibshirani(1993)}]{efron93}
Efron B., Tibshirani R., 1993, An Introduction to the Bootstrap. Chapman and
  Hall, New York, London.

\bibitem[{{Ferraro} {et~al}\mbox{.}(2006){Ferraro}, {Mucciarelli}, {Carretta},
  \& {Origlia}}]{ferraro06}
{Ferraro} F.~R., {Mucciarelli} A., {Carretta} E., {Origlia} L., 2006, \apjl,
  645, L33

\bibitem[{{Goudfrooij} {et~al}\mbox{.}(2014){Goudfrooij}, {Girardi},
  {Kozhurina-Platais}, {Kalirai}, {Platais}, {Puzia}, {Correnti}, {Bressan},
  {Chandar}, {Kerber}, {Marigo}, \& {Rubele}}]{goudfrooij14}
{Goudfrooij} P. {et~al.}, 2014, ApJ, 797, 35

\bibitem[{{Goudfrooij} {et~al}\mbox{.}(2011){Goudfrooij}, {Puzia},
  {Kozhurina-Platais}, \& {Chandar}}]{goudfrooij11}
{Goudfrooij} P., {Puzia} T.~H., {Kozhurina-Platais} V., {Chandar} R., 2011,
  \apj, 737, 3

\bibitem[{{Li}, {de Grijs} \& {Deng}(2014){Li}, {de Grijs}, \& {Deng}}]{li14}
{Li} C., {de Grijs} R., {Deng} L., 2014, Nat., 516, 367

\bibitem[{{Mackey} {et~al}\mbox{.}(2008){Mackey}, {Broby Nielsen}, {Ferguson},
  \& {Richardson}}]{mackey08}
{Mackey} A.~D., {Broby Nielsen} P., {Ferguson} A.~M.~N., {Richardson} J.~C.,
  2008, ApJL, 681, L17

\bibitem[{{Marino} {et~al}\mbox{.}(2012){Marino}, {Milone}, {Piotto},
  {Cassisi}, {D'Antona}, {Anderson}, {Aparicio}, {Bedin}, {Renzini}, \&
  {Villanova}}]{marino12}
{Marino} A.~F. {et~al.}, 2012, \apj, 746, 14

\bibitem[{{Martocchia} {et~al}\mbox{.}(2017){Martocchia}, {Bastian}, {Usher},
  {Kozhurina-Platais}, {Niederhofer}, {Cabrera-Ziri}, {Dalessandro},
  {Hollyhead}, {Kacharov}, {Lardo}, {Larsen}, {Mucciarelli}, {Platais},
  {Salaris}, {Cordero}, {Geisler}, {Hilker}, {Li}, \& {Mackey}}]{paperIII}
{Martocchia} S. {et~al.}, 2017, MNRAS, 468, 3150

\bibitem[{{Martocchia} {et~al}\mbox{.}(2018){Martocchia}, {Cabrera-Ziri},
  {Lardo}, {Dalessandro}, {Bastian}, {Kozhurina-Platais}, {Usher},
  {Niederhofer}, {Cordero}, {Geisler}, {Hollyhead}, {Kacharov}, {Larsen}, {Li},
  {Mackey}, {Hilker}, {Mucciarelli}, {Platais}, \& {Salaris}}]{martocchia17b}
{Martocchia} S. {et~al.}, 2018, \mnras, 473, 2688

\bibitem[{{Milone} {et~al}\mbox{.}(2009){Milone}, {Bedin}, {Piotto}, \&
  {Anderson}}]{milone09}
{Milone} A.~P., {Bedin} L.~R., {Piotto} G., {Anderson} J., 2009, A\&A, 497, 755

\bibitem[{{Milone} {et~al}\mbox{.}(2015){Milone}, {Bedin}, {Piotto}, {Marino},
  {Cassisi}, {Bellini}, {Jerjen}, {Pietrinferni}, {Aparicio}, \&
  {Rich}}]{milone15}
{Milone} A.~P. {et~al.}, 2015, \mnras, 450, 3750

\bibitem[{{Milone} {et~al}\mbox{.}(2017){Milone}, {Piotto}, {Renzini},
  {Marino}, {Bedin}, {Vesperini}, {D'Antona}, {Nardiello}, {Anderson}, {King},
  {Yong}, {Bellini}, {Aparicio}, {Barbuy}, {Brown}, {Cassisi}, {Ortolani},
  {Salaris}, {Sarajedini}, \& {van der Marel}}]{milone17}
{Milone} A.~P. {et~al.}, 2017, \mnras, 464, 3636

\bibitem[{{Mucciarelli} {et~al}\mbox{.}(2008){Mucciarelli}, {Carretta},
  {Origlia}, \& {Ferraro}}]{mucciarelli08}
{Mucciarelli} A., {Carretta} E., {Origlia} L., {Ferraro} F.~R., 2008, AJ, 136,
  375

\bibitem[{{Mucciarelli} {et~al}\mbox{.}(2014){Mucciarelli}, {Dalessandro},
  {Ferraro}, {Origlia}, \& {Lanzoni}}]{mucciarelli14}
{Mucciarelli} A., {Dalessandro} E., {Ferraro} F.~R., {Origlia} L., {Lanzoni}
  B., 2014, ApJ, 793, L6

\bibitem[{{Mucciarelli} {et~al}\mbox{.}(2007){Mucciarelli}, {Ferraro},
  {Origlia}, \& {Fusi Pecci}}]{mucciarelli07}
{Mucciarelli} A., {Ferraro} F.~R., {Origlia} L., {Fusi Pecci} F., 2007, AJ,
  133, 2053

\bibitem[{{Mucciarelli} {et~al}\mbox{.}(2012){Mucciarelli}, {Origlia},
  {Ferraro}, {Bellazzini}, \& {Lanzoni}}]{mucciarelli12}
{Mucciarelli} A., {Origlia} L., {Ferraro} F.~R., {Bellazzini} M., {Lanzoni} B.,
  2012, \apjl, 746, L19

\bibitem[{{Nardiello} {et~al}\mbox{.}(2015){Nardiello}, {Piotto}, {Milone},
  {Marino}, {Bedin}, {Anderson}, {Aparicio}, {Bellini}, {Cassisi}, {D'Antona},
  {Hidalgo}, {Ortolani}, {Pietrinferni}, {Renzini}, {Salaris}, {Marel}, \&
  {Vesperini}}]{nardiello15}
{Nardiello} D. {et~al.}, 2015, \mnras, 451, 312

\bibitem[{{Niederhofer} {et~al}\mbox{.}(2017{\natexlab{a}}){Niederhofer},
  {Bastian}, {Kozhurina-Platais}, {Larsen}, {Hollyhead}, {Lardo},
  {Cabrera-Ziri}, {Kacharov}, {Platais}, {Salaris}, {Cordero}, {Dalessandro},
  {Geisler}, {Hilker}, {Li}, {Mackey}, \& {Mucciarelli}}]{paperII}
{Niederhofer} F. {et~al.}, 2017{\natexlab{a}}, MNRAS, 465, 4159

\bibitem[{{Niederhofer} {et~al}\mbox{.}(2017{\natexlab{b}}){Niederhofer},
  {Bastian}, {Kozhurina-Platais}, {Larsen}, {Salaris}, {Dalessandro},
  {Mucciarelli}, {Cabrera-Ziri}, {Cordero}, {Geisler}, {Hilker}, {Hollyhead},
  {Kacharov}, {Lardo}, {Li}, {Mackey}, \& {Platais}}]{paperI}
{Niederhofer} F. {et~al.}, 2017{\natexlab{b}}, MNRAS, 464, 94

\bibitem[{{Niederhofer} {et~al}\mbox{.}(2015){Niederhofer}, {Georgy},
  {Bastian}, \& {Ekstr{\"o}m}}]{niederhofer15}
{Niederhofer} F., {Georgy} C., {Bastian} N., {Ekstr{\"o}m} S., 2015, MNRAS,
  453, 2070

\bibitem[{{Pietrinferni} {et~al}\mbox{.}(2004){Pietrinferni}, {Cassisi},
  {Salaris}, \& {Castelli}}]{pietrinferni04}
{Pietrinferni} A., {Cassisi} S., {Salaris} M., {Castelli} F., 2004, ApJ, 612,
  168

\bibitem[{{Prantzos}, {Charbonnel} \& {Iliadis}(2017){Prantzos}, {Charbonnel},
  \& {Iliadis}}]{prantzos17}
{Prantzos} N., {Charbonnel} C., {Iliadis} C., 2017, ArXiv e-prints, 1709.05819

\bibitem[{{Renzini} {et~al}\mbox{.}(2015){Renzini}, {D'Antona}, {Cassisi},
  {King}, {Milone}, {Ventura}, {Anderson}, {Bedin}, {Bellini}, {Brown},
  {Piotto}, {van der Marel}, {Barbuy}, {Dalessandro}, {Hidalgo}, {Marino},
  {Ortolani}, {Salaris}, \& {Sarajedini}}]{renzini15}
{Renzini} A. {et~al.}, 2015, \mnras, 454, 4197

\bibitem[{{Sbordone} {et~al}\mbox{.}(2011){Sbordone}, {Salaris}, {Weiss}, \&
  {Cassisi}}]{sbordone11}
{Sbordone} L., {Salaris} M., {Weiss} A., {Cassisi} S., 2011, Astron.
  Astrophys., 534, A9

\bibitem[{Scott(1992)}]{scott}
Scott D., 1992, Multivariate Density Estimation: Theory, Practice, and
  Visualization. John Wiley \& Sons, New York, Chicester

\bibitem[{{Stetson}(1987)}]{stetson87}
{Stetson} P.~B., 1987, PASP, 99, 191

\bibitem[{{Westerlund}(1997)}]{westerlund97}
{Westerlund} B.~E., 1997, Cambridge Astrophysics Series, 29

\end{thebibliography}

\label{lastpage}
\end{document}